\documentclass[a4paper,12pt]{article}
\usepackage[margin=2.5cm]{geometry}
\usepackage{graphicx}
\usepackage{epsfig}
\usepackage{amsmath}
\usepackage{amssymb}
\usepackage{cite}
\usepackage{float}
\usepackage[font={footnotesize,it}]{caption}
\usepackage{authblk}
\numberwithin{equation}{section}

\begin{document}

\begin{titlepage}

\title{Chaotic dynamics of strings in charged black hole backgrounds}
\author[1]{Pallab Basu \thanks{\noindent E-mail:~ pallab.basu@icts.res.in}}
\author[2]{Pankaj Chaturvedi \thanks{\noindent E-mail:~ cpankaj@iitk.ac.in }}
\author[3]{Prasant Samantray \thanks{\noindent E-mail:~ prasant.samantray@iiti.ac.in}}

\affil[1]{International Center for Theoretical Sciences, Tata Institute of Fundamental Research, Bangalore 560012, India}
\affil[2]{Department of Physics, Indian Institute of Technology Kanpur, Kanpur 208016, India}
\affil[3]{Centre of Astronomy, Indian Institute of Technology Indore\\ Khandwa Road, Simrol 453552, India}

\maketitle

\abstract{
\noindent
We study the motion of a string in the background of Reissner-Nordstrom black hole, in both AdS as well as asymptotically flat spacetimes. We describe the phase space of this dynamical system through largest Lyapunov exponent, Poincare sections and basins of attractions. We observe that string motion in these settings is particularly chaotic and comment on its characteristics.}

\end{titlepage}

\section{Introduction}
How strings move through spacetime is a question of central importance in string theory. Understanding how string motion differs from the motion of a particle in the cosmological or gravitational backgrounds may give us key insights and help us resolve various problems associated with classical gravitational singularities. However, in most curved backgrounds string theory is not analytically solvable. It is currently not known how to understand quantum string spectrum (even neglecting loop effects) in various curved backgrounds. Although integrability of string motion has been shown in symmetric coset spaces at classical level \cite{Mandal:2002fs,Bena:2003wd}. Using the AdS/CFT duality, it has been strongly suggested that these kind of integrability may be extended to the full theory \cite{Beisert:2010jr,Giataganas:2013dha,Giataganas:2014hma}. 

\noindent At a classical level, it was shown previously that the motion of a string on various curved backgrounds is nonintegrable, and string dynamics shows signatures of chaotic motion \cite{Basu:2011di,Basu:2011fw,Basu:2012ae,Stepanchuk:2012xi,Chervonyi:2013eja,Bai:2014wpa,Ma:2014aha,Asano:2015qwa,Panigrahi:2016zny}. In contrast, motion of a test particle in most of these radially symmetric backgrounds is integrable, implying - a string due to its extended nature may show a more complex behavior. Aspects of string dynamics in the background of Schwarzchild black hole has been studied in \cite{Frolov:1999pj,Zayas:2010fs}. In this paper we would review the same, and also look at the interesting problem of classical string motion in Reissner Nordstrom (RN) black hole backgrounds. Other than the chaotic aspects of motion, there are many fold motivations to study string dynamics in a black hole background. For example, scattering of cosmic strings by primordial black holes may have effect on the density perturbations in the early epoch. Additionally, it is conjectured that during the inflationary era, not only longer cosmic strings were preferred entropically, but also the expansion itself might have produced strings of astronomical proportions. A cosmic string of the order of a megaparsec would have observationally testable signatures like lensing etc. The case for a string in Reissner Nordstrom (RN) black hole is also special since it possesses an extremal limit. The extremal RN black hole is interesting because its near horizon limit is $AdS_2 \times S^2$, which can qualitatively change the string dynamics leading to interesting features. Extremality is one of the key ingredients necessary for the successful microscopic counting of black hole entropy in string theory, as it allows us more control of the theory. Since we are interested in the chaotic dynamics, it is but natural to investigate how this ``determinism" is affected as we move away from extremality.

In this paper, we model a circular classical string in the coaxial configuration in presence of a black hole. Unlike the point particle, a string would have vibrational degrees of freedom and is free to vibrate/oscillate in its plane. This is a key difference and vastly differentiates the dynamics of a point particle and a string leading to interesting consequences. In our setup the string propagates in the direction perpendicular to its plane, and therefore axial symmetry is manifest. As we shall observe, this simple and symmetric set up captures the essential features required to understand string dynamics in non-trivial charged black hole backgrounds.

\section{Ring string in (3+1) dimensional charged black hole background}
To describe the dynamics of a string in (3+1) dimensional charged black hole backgrounds we begin with the Polyakov action which governs the dynamics of string
\begin{equation}
S = \frac{-1}{2\pi\alpha'} \int \sqrt{-h}h^{ab} G_{\mu\nu}\partial_a X^\mu \partial_b X^\nu d\tau d\sigma, \label{PolyakovS}
\end{equation} 
with $h^{ab}$ being the worldsheet metric and $G_{\mu\nu}$ the target space metric. $X^\mu$ are the target space coordinates and the indices $(\{a,b\} = 1,2)$ stand for $(\tau,\sigma)$ which we denote as the coordinates on the worldsheet of the string. Our target space metric  $G_{\mu\nu}$ describes a charged black hole background and may be taken as
\begin{equation}
ds^2 = -f(r)dt^2 + \frac{dr^2}{f(r)} + r^2 \left(d\theta^2 + \sin^2\theta d\phi^2 \right), \label{Tmetric}
\end{equation}
where, $f(r) = \left(1 - \frac{2M}{r} + \frac{Q^2}{r^2}\right)$ for  Reissner-Nordstrom (RN) blackhole in asymptotically flat space-time, and $f(r)=\left(1 - \frac{2M}{r} + \frac{Q^2}{r^2} +\frac{r^2}{l^2}\right)$ for the Reissner-Nordstrom (RN) black hole in $AdS_4$ spacetime, with $l$ as the $AdS$ length. 
\begin{figure}[H]
\centering
\includegraphics[width =3.5in,height=2.2in]{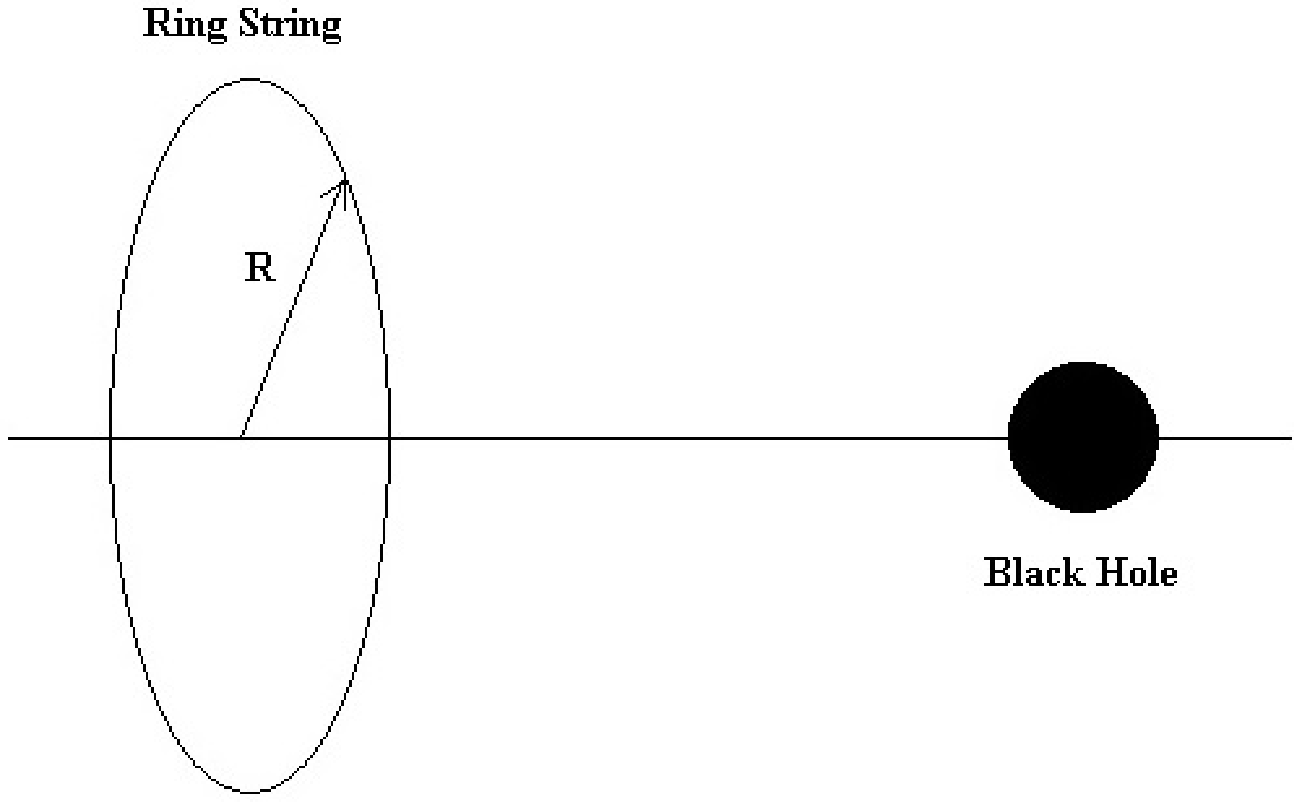}
\caption{\label{fig:string} }\end{figure}

The circular string configuration of our interest drawn in fig.(\ref{fig:string}), is described by the following embedding in the target space
\begin{equation}
t=t(\tau),~~~r = r(\tau),~~~\theta = \theta(\tau),~~~\phi = n \sigma,\label{ansatz}
\end{equation}
where ``$n$" is the winding number of the string along the $\phi$ dimension. Substituting the above ansatz in (\ref{PolyakovS}) in the conformal gauge $h^{ab} = \eta^{ab}$, we obtain the following form of the Lagrangian 
\begin{eqnarray}
{\cal L} &=& \frac{-1}{2\pi\alpha'} \left[f\dot{t}^2 - \frac{\dot{r}^2}{f} - r^2 \dot{\theta}^2 + r^2  n^2 \sin^2 \theta \right], \label{PolyakovL}
\end{eqnarray}
with the equations of motion as
\begin{eqnarray}
\ddot{r}&=&\frac{ f'}{2 f}\dot{r}^2-\frac{f f'}{2}  \dot{t}^2 + r f \left(\dot{\theta }^2-n^2 \sin ^2\theta\right),\label{reom}\\
\ddot{\theta}&=&-\frac{2 }{r}\dot{\theta} \dot{r}-n^2 \sin \theta \cos\theta, ~~~~\dot{t}=\frac{E_n}{f},\label{theta&teom}
\end{eqnarray}
here, ``$E_n$" is related to the energy of the string which is a constant of motion. Dot superscript implies differentiation with respect to $\tau$, and the prime represents derivative with respect to $r$. The conformal gauge $h^{ab} = \eta^{ab}$ yields the following non-trivial constraint
\begin{equation}
G_{\mu\nu}\left(\partial_{0}X^{\mu}\partial_{0}X^{\nu}+\partial_{1}X^{\mu}\partial_{1}X^{\nu}\right)=0,\label{cons1}
\end{equation}
and in the background given by eq.(\ref{Tmetric}) with eq.(\ref{theta&teom}), it takes the following form
\begin{equation}
\dot{r}^2+r^2f\dot{\theta}^2+n^2r^2f\sin^2\theta=E_n^2,\label{cons2}
\end{equation}

Working out the Hamiltonian equations of motion for the Lagrangian (\ref{PolyakovL}), we get the following set of coupled first-order linear differential equations.
\begin{eqnarray}
\dot{t} &=& \frac{-\pi\alpha'}{f} P_t,\label{Hteom} \\
\dot{P_t} &=& 0 , \label{Hpteom}  \\
\dot{r} &=& \pi\alpha' f P_r, \label{Hreom} \\
\dot{P_r} &=& \frac{-\pi\alpha' f' P^2_r}{2} - \frac{\pi\alpha' f' P^2_t}{f^2} + \frac{\pi\alpha' P^2_{\theta}}{r^3} - \frac{n^2 r \sin^2 \theta}{\pi\alpha'}, \label{Hpreom}  \\
\dot{\theta} &=& \frac{\pi\alpha P_{\theta}}{r^2}, \label{Hthetaeom}  \\
\dot{P}_{\theta} &=& \frac{-n^2 r^2 \sin \theta \cos\theta}{\pi\alpha'} , \label{Hpthetaeom} \\
{\cal H} &=& \frac{\pi\alpha'}{2} \left[f P^2_r + \frac{P^2_{\theta}}{r^2} -  \frac{P^2_{t}}{f^2}\right]  + \frac{n^2 r^2 \sin^2\theta}{2\pi\alpha'}, \label{Hamil} 
\end{eqnarray}  
where $f' = \frac{df(r)}{dr}$. We now set the Hamiltonian constraint as ${\cal H}=0$. This comes from fixing the conformal gauge resulting in the equation $T_{ab}=0$. Additionally, we also set $P_t = E_n$ as the conserved energy, which is also the conserved world sheet charge.  

The constraint given by eq.(\ref{cons2}) represents the energy conservation condition for the motion of circular ring in the charged black hole background. It is to be noted that the various possible modes of motion for the ring string are characterized by the solutions $r(\tau)$, $\theta(\tau)$ and the radius $R(\tau)=r\sin\theta$ of the string. Following \cite{Frolov:1999pj}, the ring string in the AdS or asymptotically flat RN background can be anticipated to have following modes of motion 
\begin{itemize}
\item  Ring string in the asymptotically flat charged black hole background
\begin{enumerate}
\item[i)]   The ring string can fly past the black hole and escape to infinity, $r\rightarrow \infty$.
\item[ii)]  The ring string can get back scattered by the black hole before escaping to infinity, $r\rightarrow \infty$.
\item[iii)] The ring string can get captured by falling into the black hole horizon $r\leq 2M$.  
\end{enumerate}
\item  Ring string in asymptotically AdS charged black hole background
\begin{enumerate}
\item[i)]   The ring string oscillates back and forth around the black hole till eternity.
\item[ii)]  The ring string completes a finite number of oscillations around the black hole before being captured by it.
\item[iii)] The ring string completes a finite number of oscillations around the black hole before escaping to infinity, $r\rightarrow \infty$.
\end{enumerate}
\end{itemize}

If the radius $R$ of the string is very small and $P_\theta=0$ to begin with, then we expect the string to behave like a point particle. In this limit, the `$\theta$' and `$r$' degrees of freedom decouple, with initial momentum in radial direction determining if the string would get captured or escape to infinity. Away from this limit we should of course expect deviations from point particle behavior. To see this, we start with an initial configuration such that the energy in `$\theta$' direction (both kinetic and potential) is more than certain critical threshold. We now embark upon numerically studying the dynamical system.  

\begin{figure}[H]
\centering
\begin{minipage}[b]{0.5\linewidth}
\includegraphics[width =3.3in,height=1.7in]{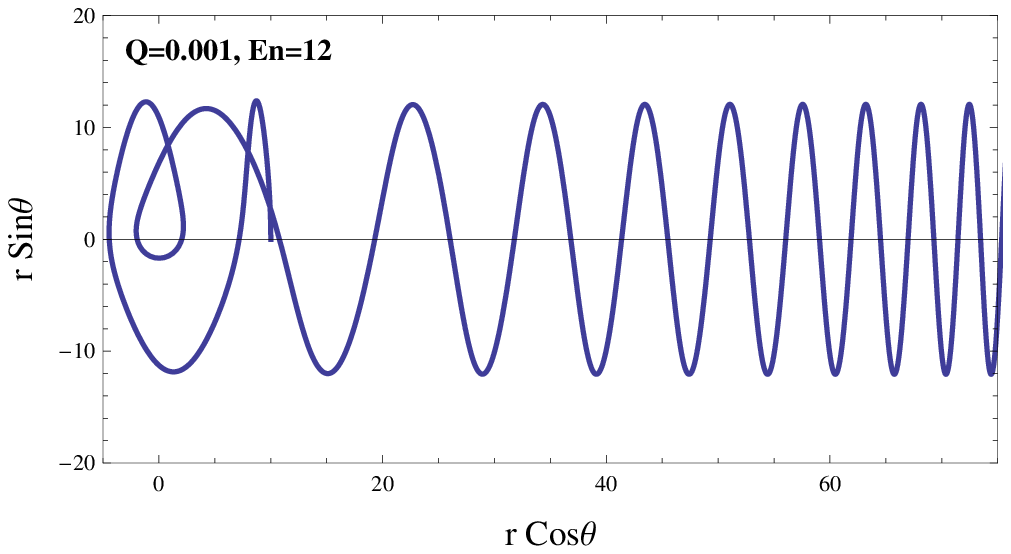}
\end{minipage}\\
\begin{minipage}[b]{0.5\linewidth}
\includegraphics[width =3.3in,height=1.7in]{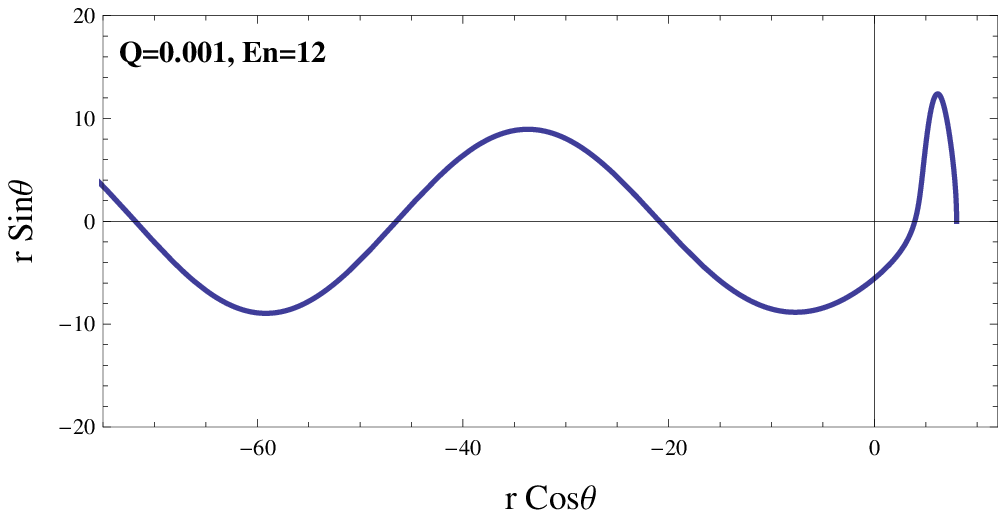}
\end{minipage}\\
\begin{minipage}[b]{0.5\linewidth}
\includegraphics[width =3.3in,height=1.7in]{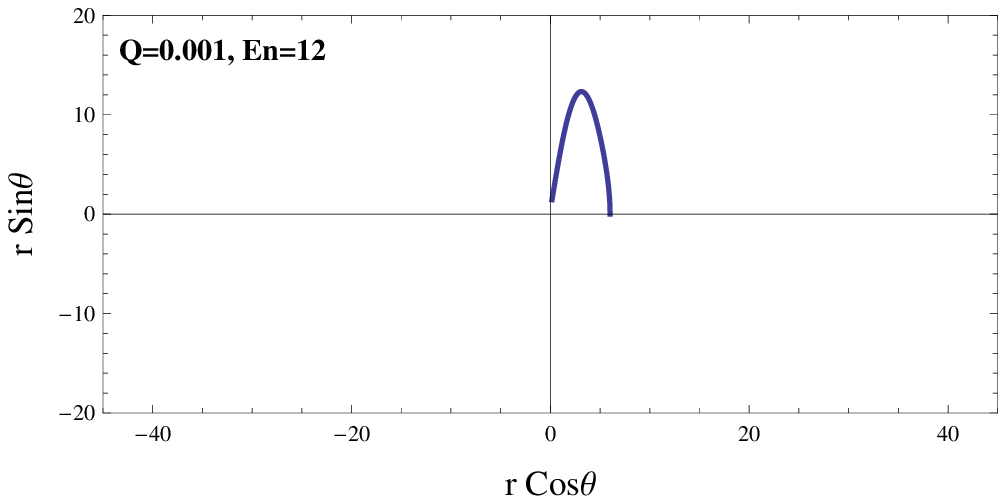}
\end{minipage}%
\caption{\label{fig:Q001En12ARN} Plots showing the  ring string trajectory in asymptotically flat RN black hole background for $Q=0.001$ with the initial conditions  $E_n=12,~\theta=0,~\frac{d}{d\tau}(r\cos{\theta})=0$ and $r=(10,8,6)$ respectively from top to bottom. The  top plot shows the backscattering of the string, the middle plot shows the escape to infinity and the bottom plot shows the capture of string by the black hole.}
\end{figure}

\section{Numerical study of the dynamical system}
In this section we describe the numerical techniques to solve the equations of motion of a ring string in charged black hole backgrounds. For this purpose, we use fourth order Runge-Kutta scheme which ensures a robust control on the error propagation and accuracy. The accuracy of the solution obtained through the numerical computation is assessed by the constraint $|H|<\delta$ for any given time, where $\delta$ is the error tolerance defined by the ``AccuracyGoal" command in the NDSolve routine. It is to be noted that our numerical scheme is not sensitive to the step size $\Delta \tau$ in time and thus provides a robust accuracy in computations.  

\subsection{Ring string in asymptotically flat RN black hole background}
\begin{figure}[H]
\centering
\begin{minipage}[b]{0.5\linewidth}
\includegraphics[width =2.5in,height=1.5in]{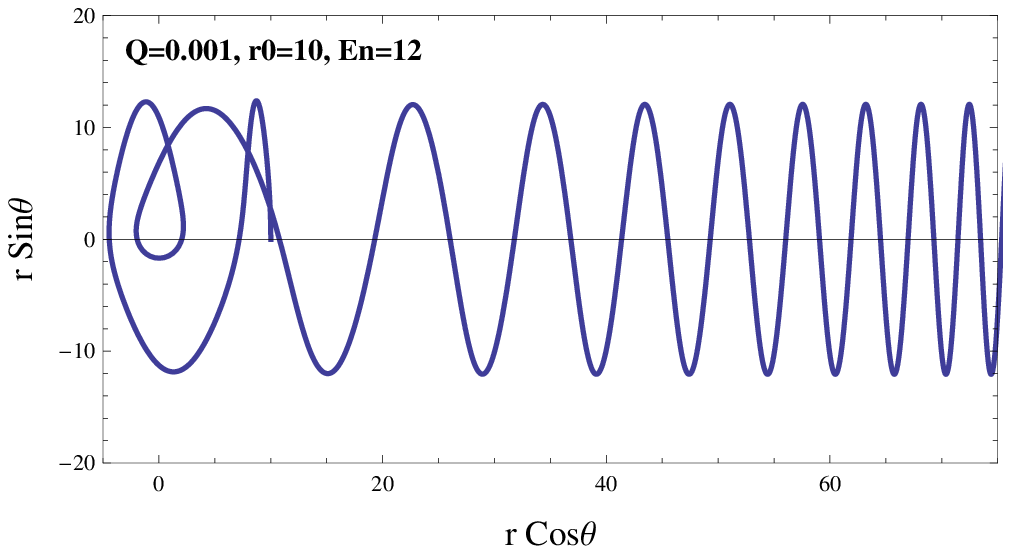}
\end{minipage}%
\begin{minipage}[b]{0.5\linewidth}
\includegraphics[width =2.5in,height=1.5in]{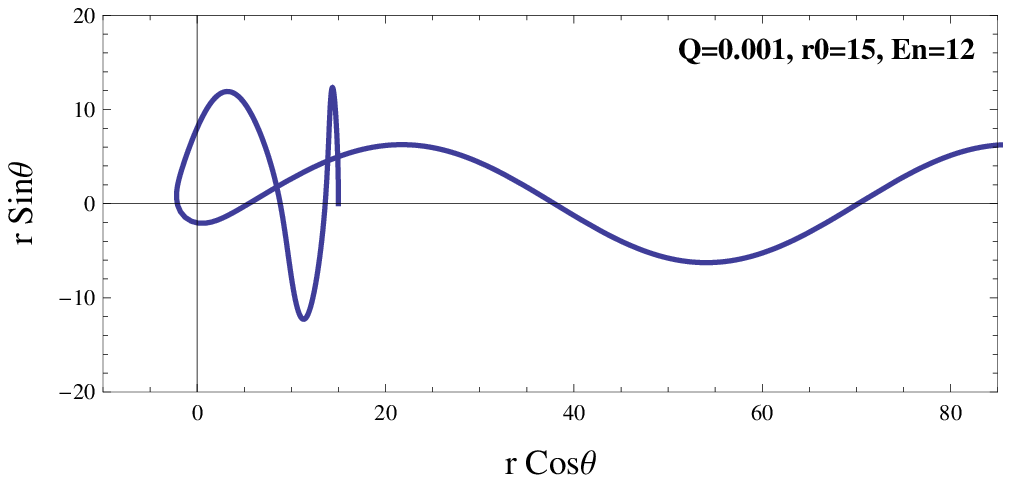}
\end{minipage}\quad
\begin{minipage}[b]{0.5\linewidth}
\includegraphics[width =2.5in,height=1.5in]{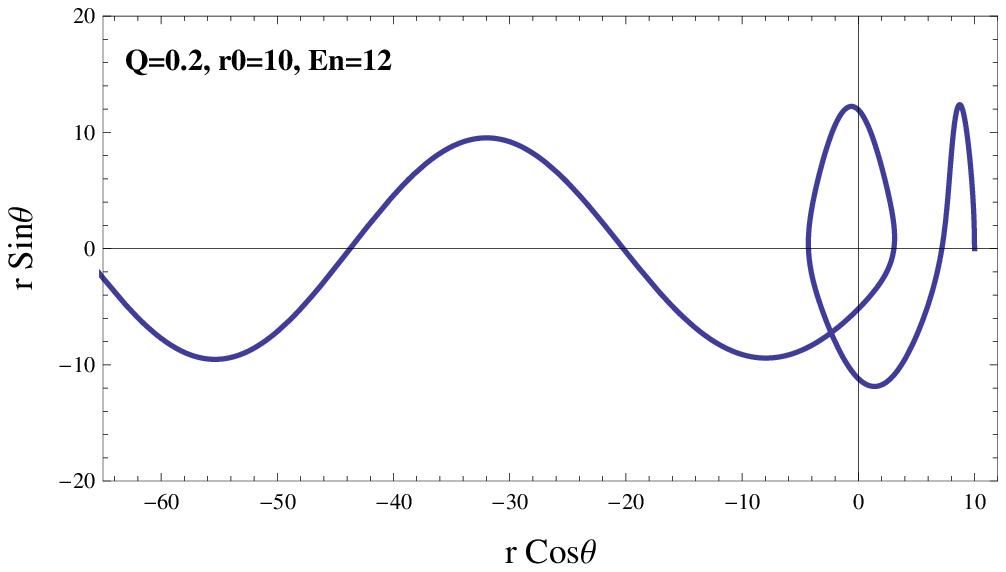}
\end{minipage}%
\begin{minipage}[b]{0.5\linewidth}
\includegraphics[width =2.5in,height=1.5in]{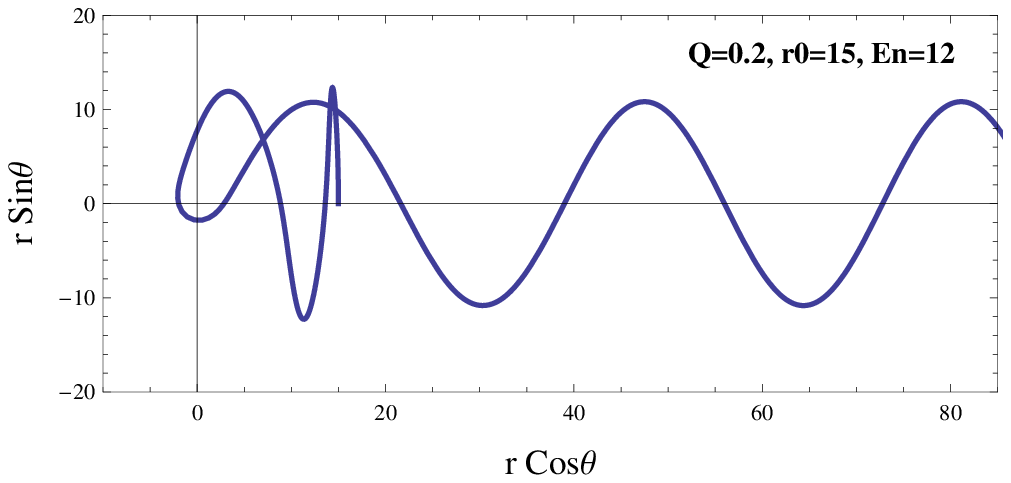}
\end{minipage}\quad
\begin{minipage}[b]{0.5\linewidth}
\includegraphics[width =2.5in,height=1.5in]{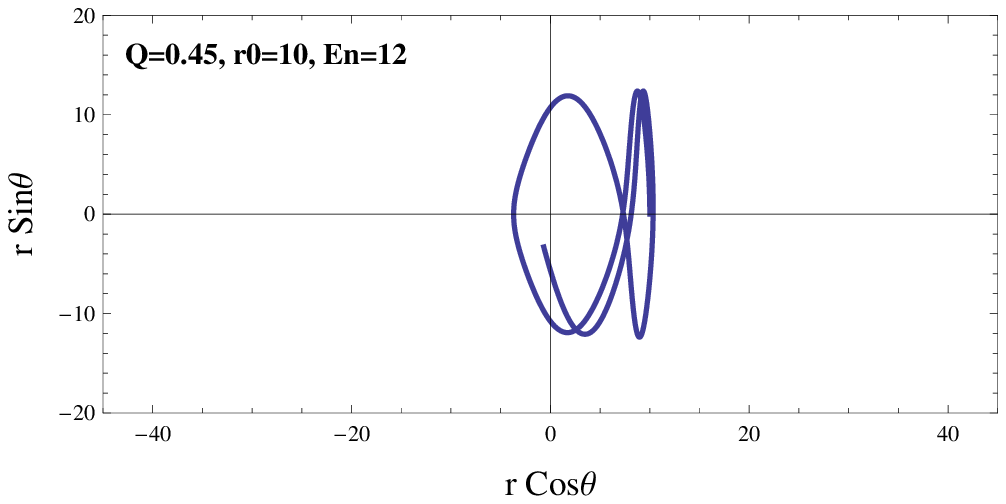}
\end{minipage}%
\begin{minipage}[b]{0.5\linewidth}
\includegraphics[width =2.5in,height=1.5in]{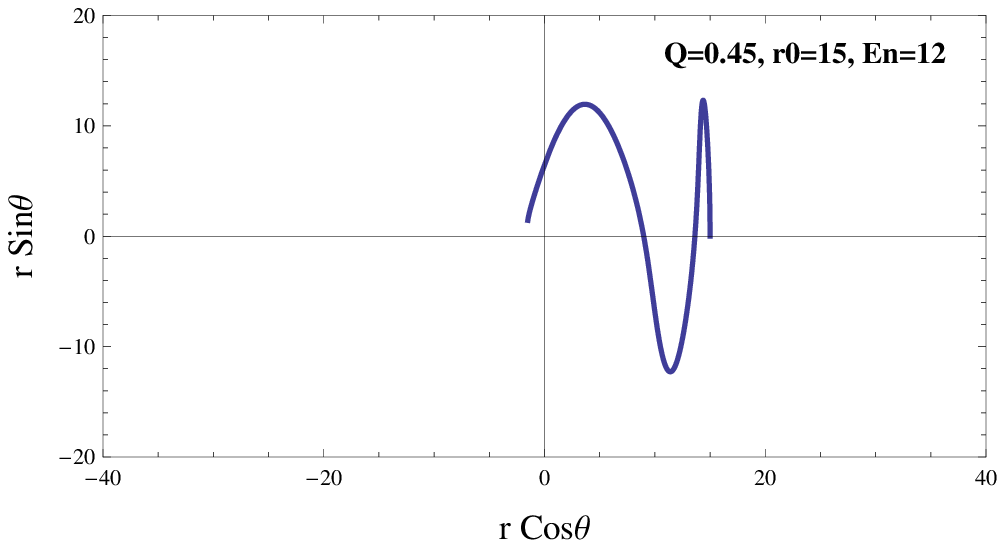}
\end{minipage}%
\caption{\label{fig:En12ARN} Plots showing the  ring string trajectory in asymptotically flat RN black hole background for $Q=(0.001,0.2,0.45)$ with the initial condition $E_n=12,~\theta=0,~\frac{d}{d\tau}(r\cos{\theta})=0,~r=10$ on the left panel and with the initial condition $E_n=12,~\theta=0,~\frac{d}{d\tau}(r\cos{\theta})=0,~r=15$ on the right panel of the figure.  }
\end{figure}

As described earlier, the ring string in the asymptotically flat charged black hole background follows three asymptotic modes of motion involving either falling in to the black hole, escaping to infinity or escaping to infinity via backscattering. Besides these three modes, there also exists an infinite set of unstable periodic orbits. In  figures (\ref{fig:Q001En12ARN}) and (\ref{fig:En12ARN}) we show these three asymptotic modes, as well as several unstable periodic modes of motion of the ring string for different values of charge $(Q)$ of the black hole.  The ring string trajectories shown in figures (\ref{fig:Q001En12ARN}) and (\ref{fig:En12ARN}) correspond to the following  initial conditions 
\begin{eqnarray}
E_n=12,~ M=1/2,~\alpha=1/\pi,~n=1,~\theta(\tau)=0,~\frac{d}{d\tau}(r\cos{\theta})=0~~at~~\tau=0.\label{iniARN}
\end{eqnarray}

As shown in figures (\ref{fig:Q001En12ARN}) and (\ref{fig:En12ARN}), there are unstable periodic orbits which demonstrate that the ring string starting a little further away from the black hole oscillates a number of times before reaching it \cite{Frolov:1999pj}. Furthermore, from these plots we observe that as we increase the black hole charge, the tendency of the string being captured increases, whereas the tendency of the string escaping to infinity decreases for the same set of initial conditions.  

\begin{figure}[H]
\centering
\begin{minipage}[b]{0.5\linewidth}
\includegraphics[width =2.7in,height=2in]{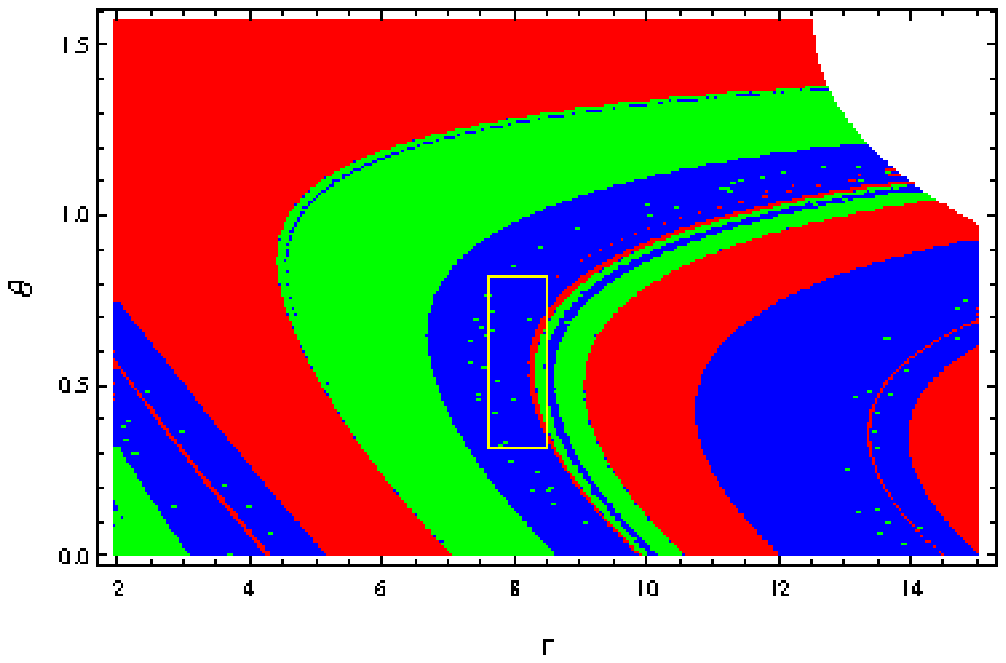}
\end{minipage}%
\begin{minipage}[b]{0.5\linewidth}
\includegraphics[width =2.7in,height=2in]{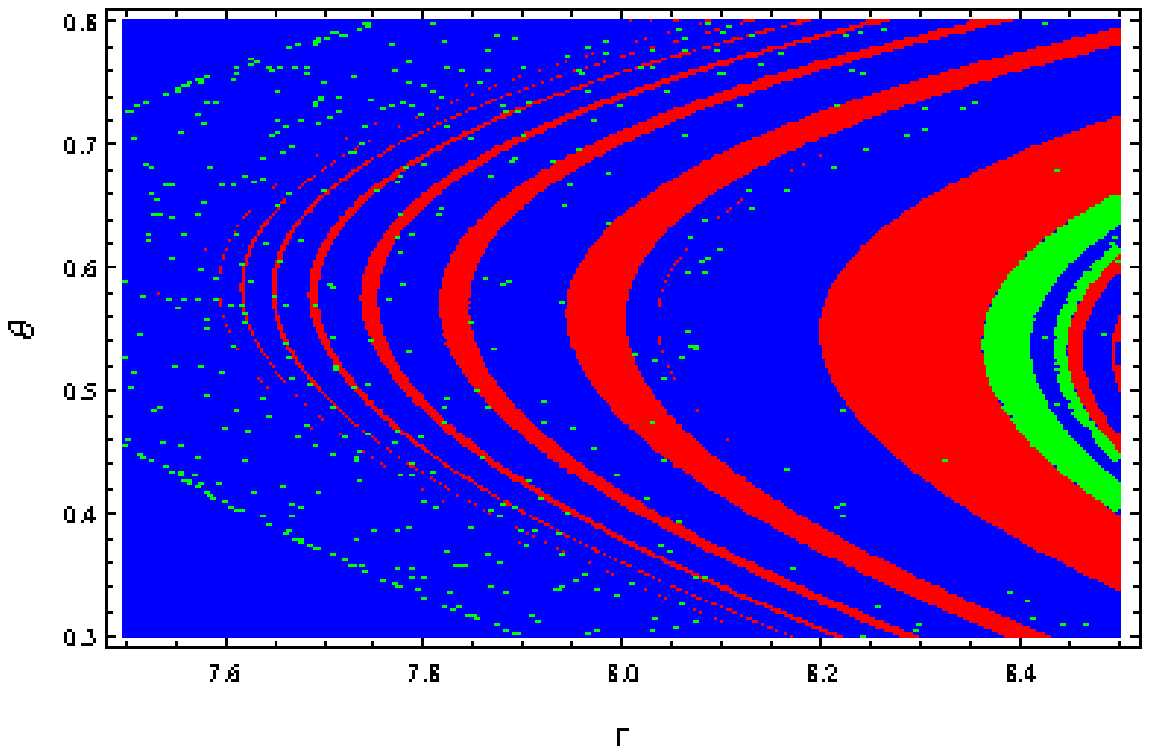}
\end{minipage}\quad
\begin{minipage}[b]{0.5\linewidth}
\includegraphics[width =2.7in,height=2in]{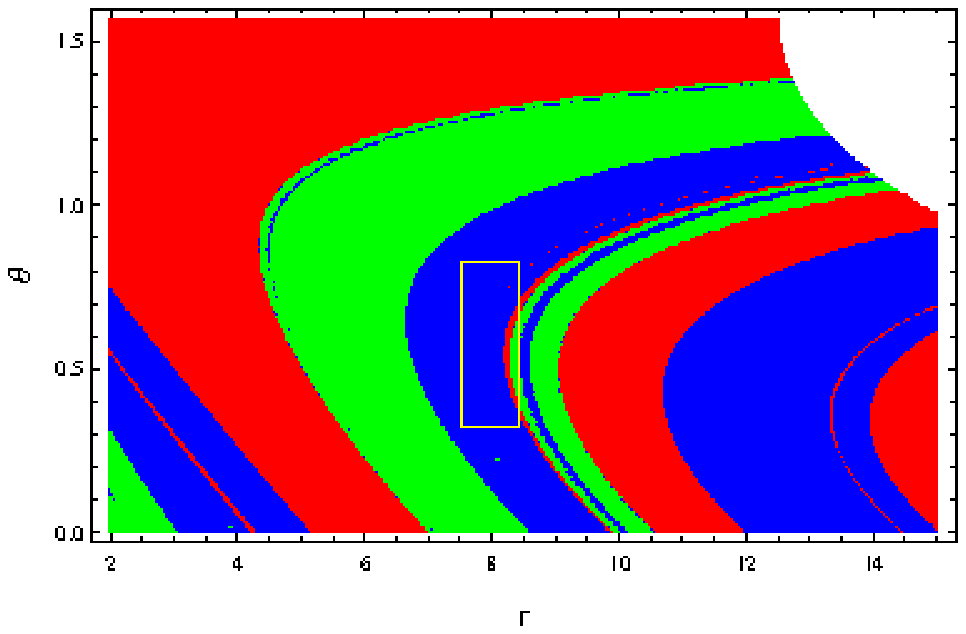}
\end{minipage}%
\begin{minipage}[b]{0.5\linewidth}
\includegraphics[width =2.7in,height=2in]{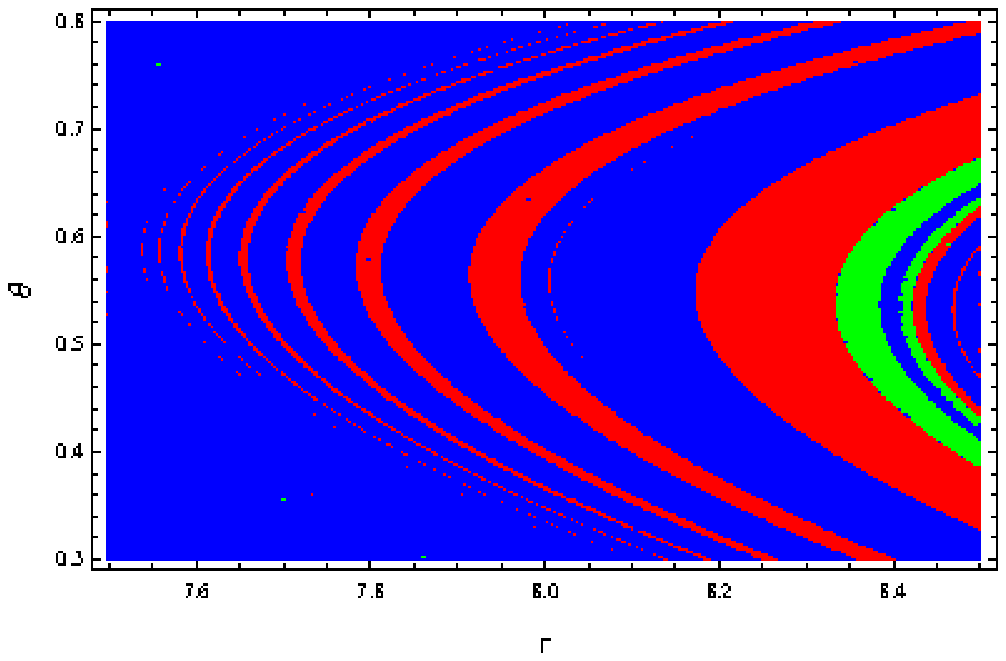}
\end{minipage}\quad
\caption{\label{fig:BasinARNQ001} Basin boundaries, plotted for different values of the charge of the black hole $Q=0.001$ (top) and $Q=0.2$ (bottom) with the initial conditions $E_n=12$, $\frac{d}{d\tau}(r\cos{\theta})=0$ with $r\in[2,17]$ on the horizontal axis and $\theta\in[0,\pi/2]$ on the vertical axis. The figure on the right side shows a detailed section of the basin boundary diagram shown on the left.}
\end{figure}

\begin{figure}[H]
\centering
\begin{minipage}[b]{0.5\linewidth}
\includegraphics[width =2.7in,height=2in]{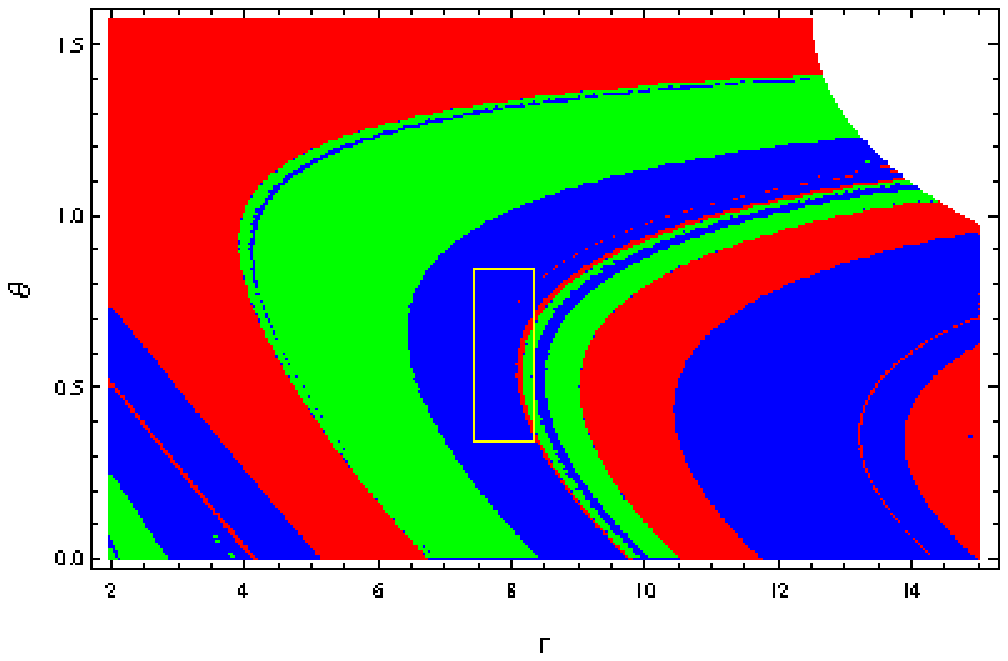}
\end{minipage}%
\begin{minipage}[b]{0.5\linewidth}
\includegraphics[width =2.7in,height=2in]{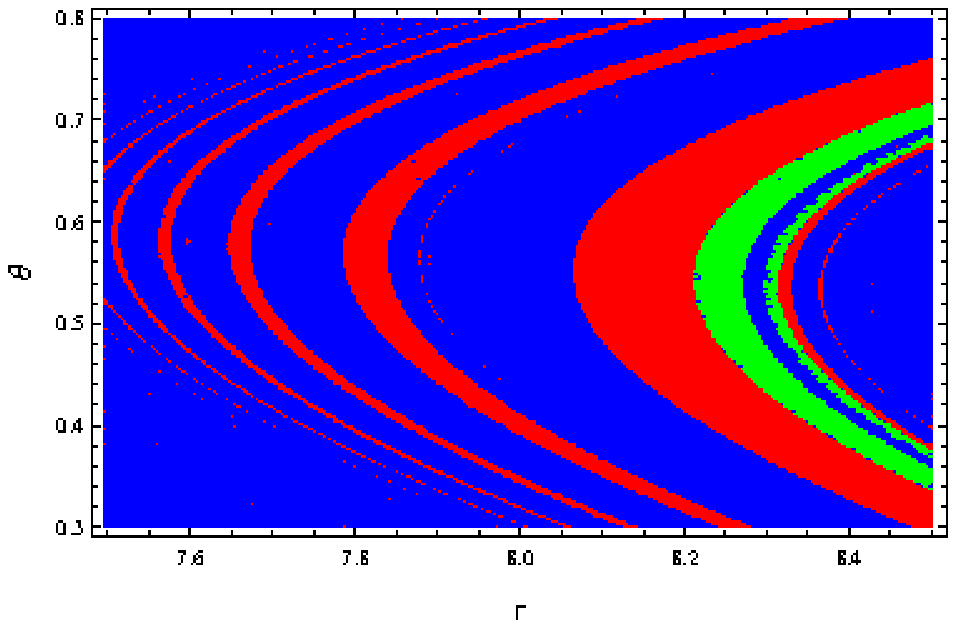}
\end{minipage}\quad
\begin{minipage}[b]{0.5\linewidth}
\includegraphics[width =2.7in,height=2in]{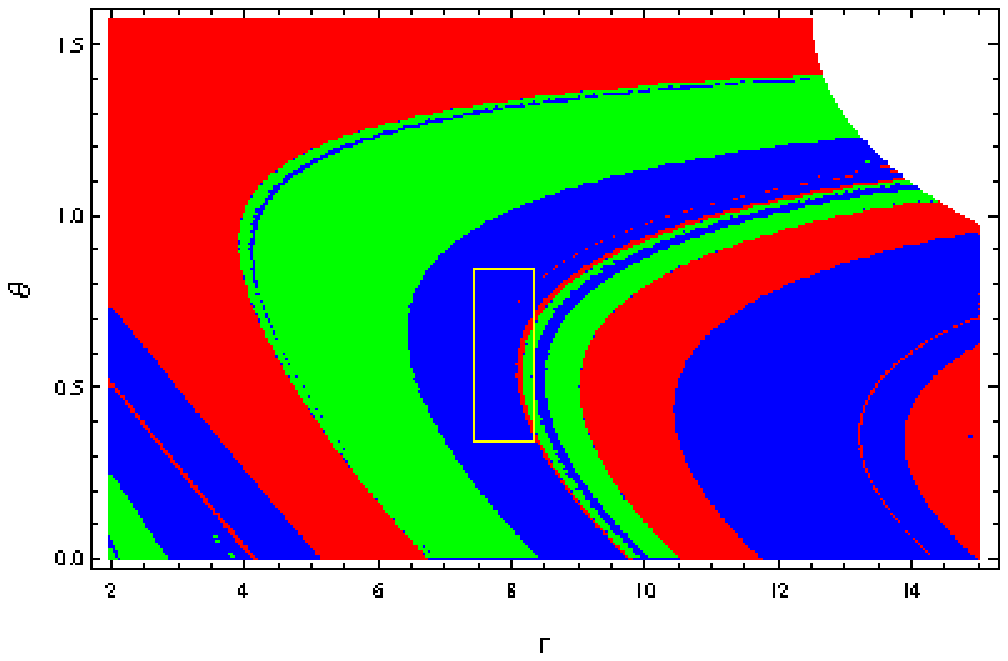}
\end{minipage}%
\begin{minipage}[b]{0.5\linewidth}
\includegraphics[width =2.7in,height=2in]{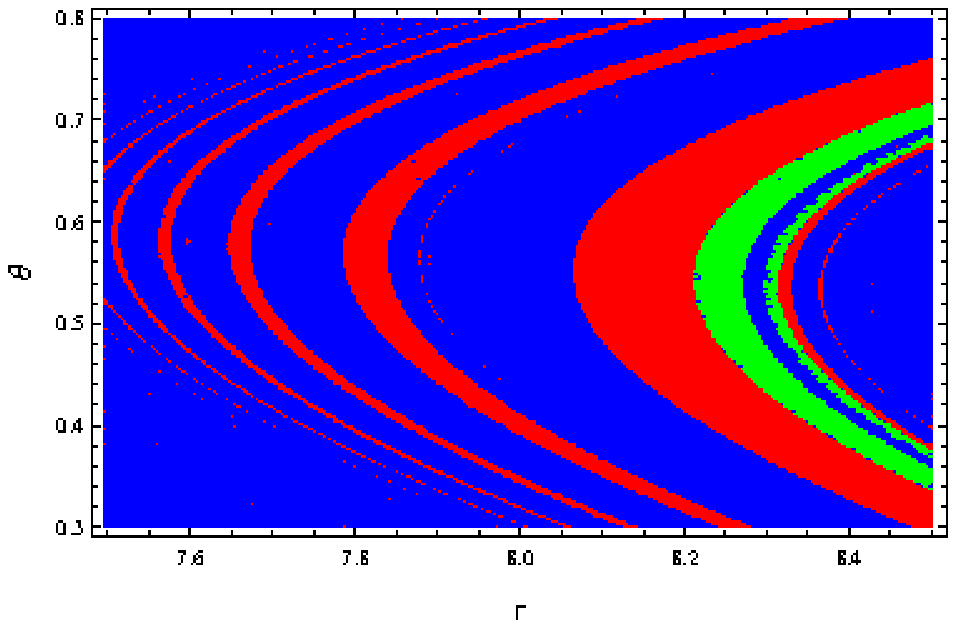}
\end{minipage}\quad
\caption{\label{fig:BasinARNQ45}  Basin boundaries, plotted for the initial conditions $Q=0.45$ (top) and $Q=0.4987$ (extremal value, bottom) with the initial conditions $E_n=12$, $\frac{d}{d\tau}(r\cos{\theta})=0$ with $r\in[2,17]$ on the horizontal axis and $\theta\in[0,\pi/2]$ on the vertical axis.The figure on the right side shows a detailed section of the basin boundary diagram shown on the left.}
\end{figure}

Additionally, in order to get a better perspective of the string dynamics in asymptotically flat RN black hole background, we adopt the basin-boundary method for coloring the two dimensional slice $(r,\theta)$ of the four dimensional phase space $(r,P_r,\theta, P_\theta)$. The two dimensional slice of the initial conditions in $(r,\theta)$ is obtained by considering the constraint given by the equation (\ref{iniARN}), which fixes the values of $(P_r, P_\theta)$ in terms of $r,\theta$, energy $E_n$ of the string, and the charge $(Q)$ of the black hole. Figures \ref{fig:BasinARNQ001} and \ref{fig:BasinARNQ45} depict the basins of attraction of the ring string for different values $Q=0.001,0.2,0.45$ and $Q=0.50$ (extremal value) of the charge of the asymptotically flat RN black hole. We have colored the $(r,\theta)$ slice of initial conditions according to different asymptotic modes of motion that the ring string demonstrates. The red colored region in figures \ref{fig:BasinARNQ001} and \ref{fig:BasinARNQ45} correspond to the case when the ring string passes beyond $(r(\tau)<r_h)$ the radius of horizon $(r_h)$ of the black hole and is captured  by it. The blue region corresponds to the case for the ring string getting scattered by the black hole before escaping to infinity $(r\rightarrow\infty)$, and the green region depicts the case when the ring string flies past the black hole and eventually escapes to infinity. Due to numerical reasons, instead of infinity the condition for escape is considered to be at some large but finite value of $r$ which we take to be $r\geq 200 r_h$. In principle the string may cross this cutoff in one direction and return back, however this will lead to a wrong color for very few points which also follows from the asymptotic behavior of potential in the Hamiltonian given by eq.(\ref{Hamil}). 
From the basin-boundary diagrams shown in figures (\ref{fig:BasinARNQ001}) and (\ref{fig:BasinARNQ45}) it is clear that there are well defined red, blue and green regions separated from each other by fuzzy boundaries. In order to verify this we magnify the region bounded by yellow box in the figures (\ref{fig:BasinARNQ001}) and (\ref{fig:BasinARNQ45}). Subsequently it is observed that the green region tends to disappear as we increase the charge of the black hole. This once again implies that the tendency of the string escaping to infinity decreases with the increasing values of the black hole charge. The magnification of the basin-boundary region bounded by yellow box also reveals a fractal structure characteristic to the systems following deterministic chaos - indicating that the dynamics of the ring string in the asymptotically flat RN black hole background is indeed chaotic. 

\subsection{Ring string in  asymptotically AdS Reissner-Nordstrom black hole background}
In this section we describe the dynamics of a ring string in the asymptotically AdS Reissner-Nordstrom (RN) black hole background. Here the three asymptotic modes of string motion involve (i) falling in to the black hole, (ii) oscillating a finite number of times before falling into the black hole, and (iii) escaping to infinity after completing some finite number of oscillations. Again, our goal is to study the string dynamics as a function of the black hole charge, in particular, the near extremal limit. We fix the parameters of our system in RN-AdS background as,
\begin{eqnarray}
E_n=12,~ M=1/2,~\alpha=1/\pi,~n=1,~l=15.\label{iniRNAdS}
\end{eqnarray}

\begin{figure}[H]
\centering
\begin{minipage}[b]{0.35\linewidth}
\includegraphics[width =2in,height=1.6in]{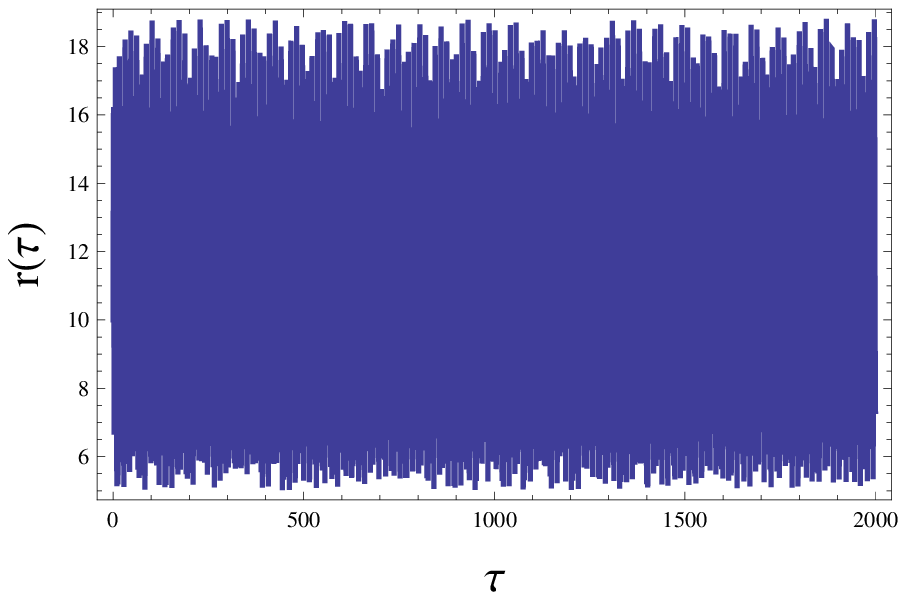}
\end{minipage}%
\begin{minipage}[b]{0.35\linewidth}
\includegraphics[width =2in,height=1.6in]{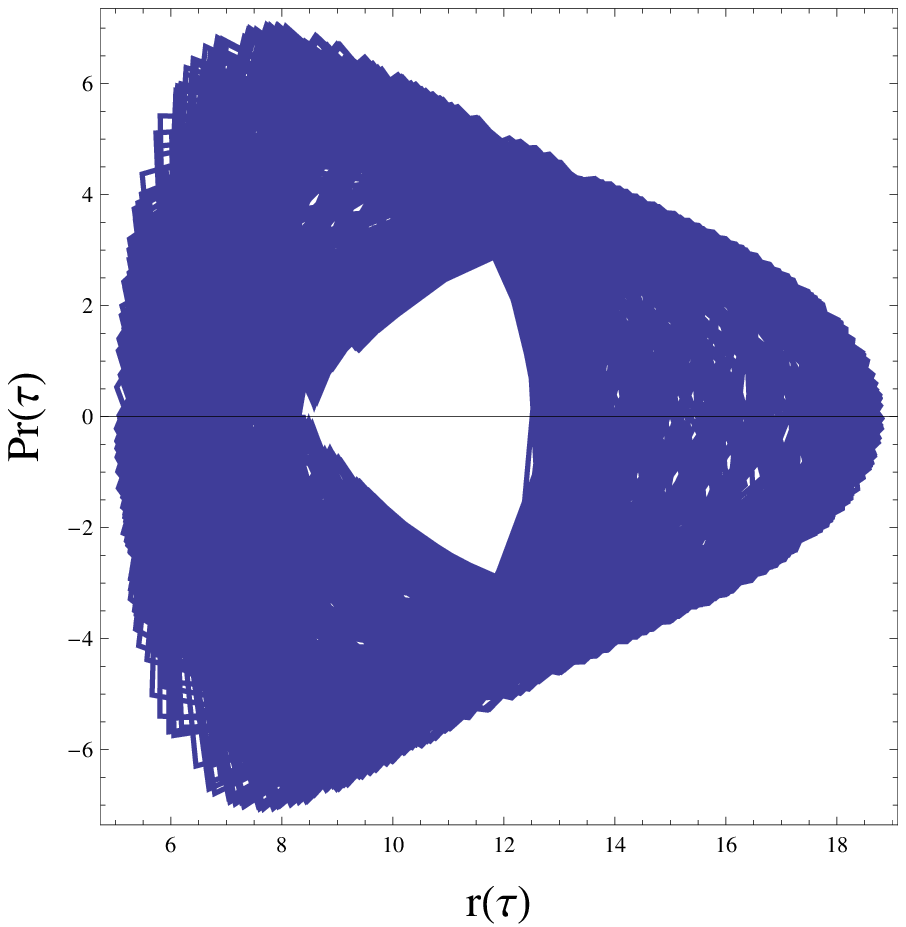}
\end{minipage}%
\begin{minipage}[b]{0.35\linewidth}
\includegraphics[width =2in,height=1.6in]{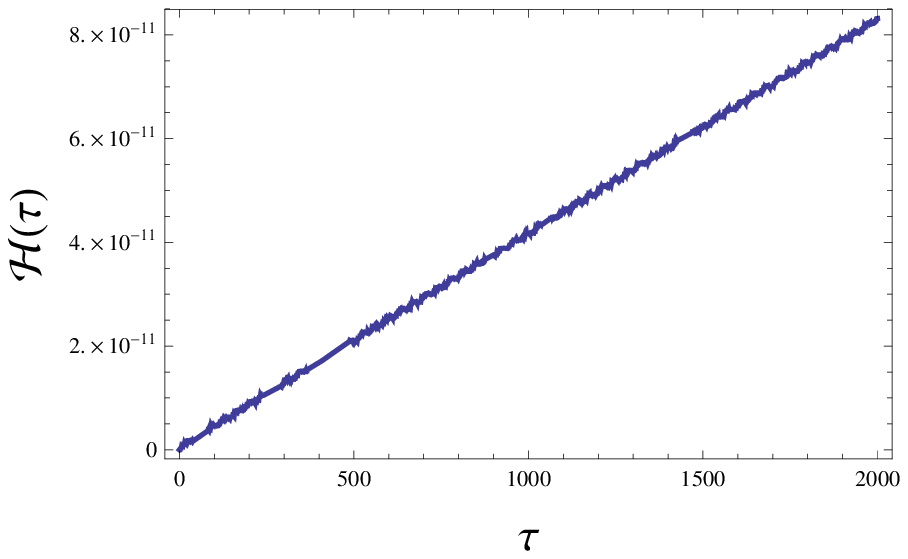}
\end{minipage}\quad
\begin{minipage}[b]{0.35\linewidth}
\includegraphics[width =2in,height=1.6in]{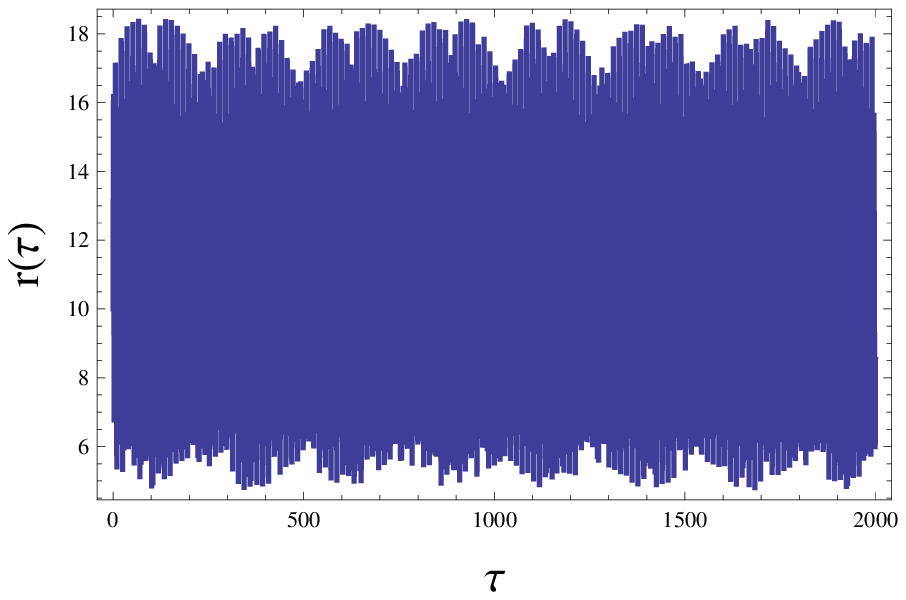}
\end{minipage}%
\begin{minipage}[b]{0.35\linewidth}
\includegraphics[width =2in,height=1.6in]{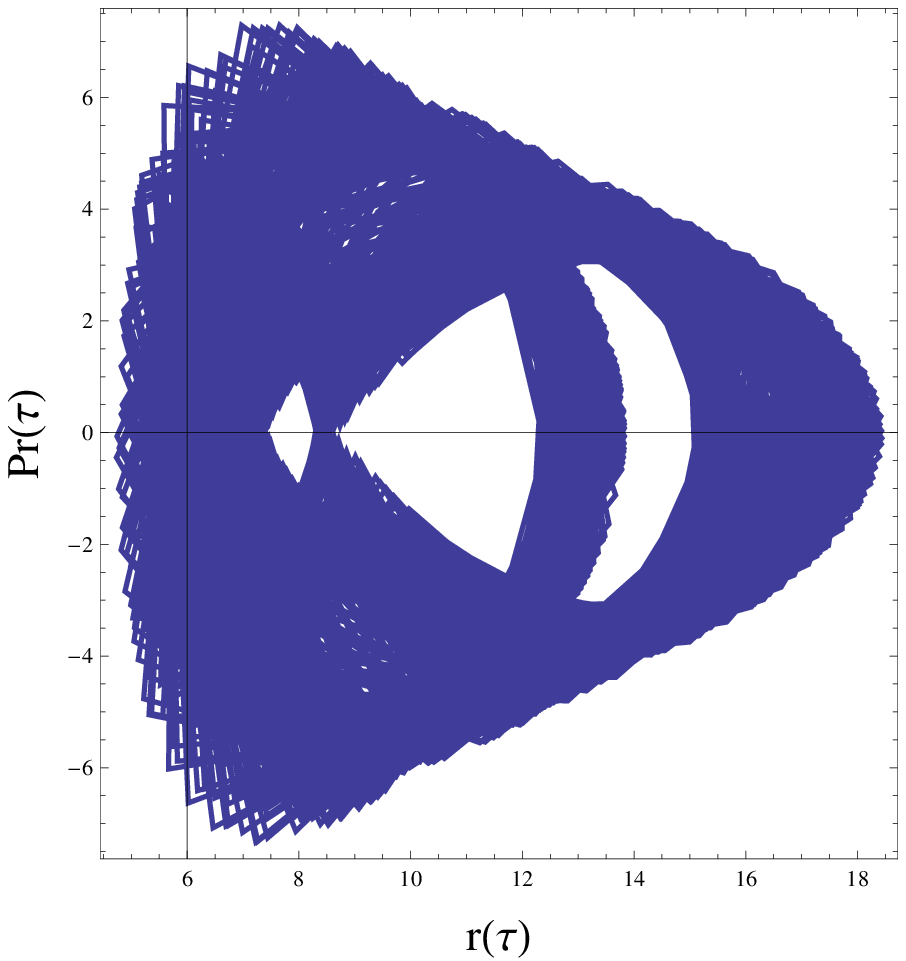}
\end{minipage}%
\begin{minipage}[b]{0.35\linewidth}
\includegraphics[width =2in,height=1.6in]{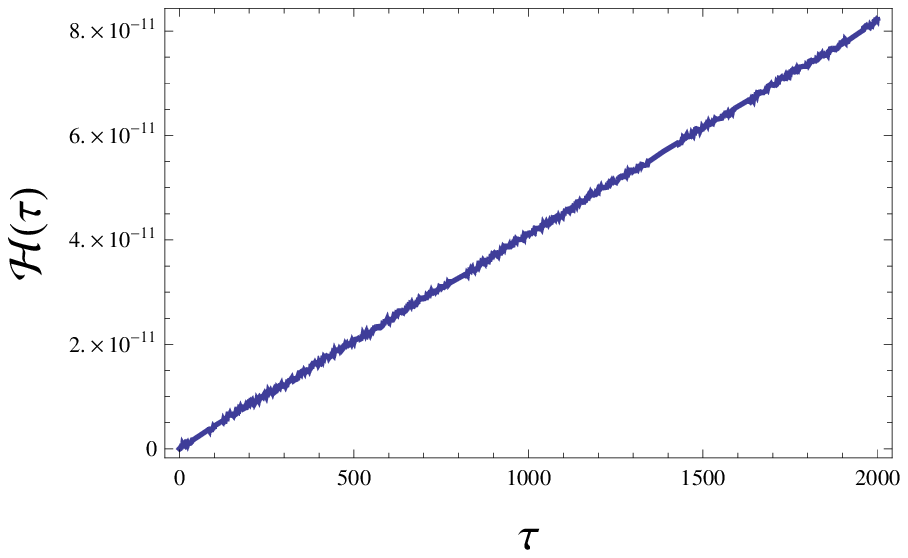}
\end{minipage}\quad
\caption{\label{fig:rvtRNAdS}  Plots from top to bottom depict the solution $r(\tau)$ and phase space plots for the ring string in RN-AdS black hole background for increasing values of the charge $Q=0.001$ and $Q=0.45$ of the black respectively,with the initial conditions $E_n=12,P_r=3.6,~r=10$ and $\theta=0$. The Hamiltonian $({\cal H})$ as a function of proper time $\tau$  (with the error tolerance $\delta= 10^{-10}$) for the corresponding  solutions $r(\tau)$ is plotted in the right panel of the figure. }
\end{figure}

\begin{figure}[H]
\centering
\begin{minipage}[b]{0.5\linewidth}
\includegraphics[width =2.7in,height=2in]{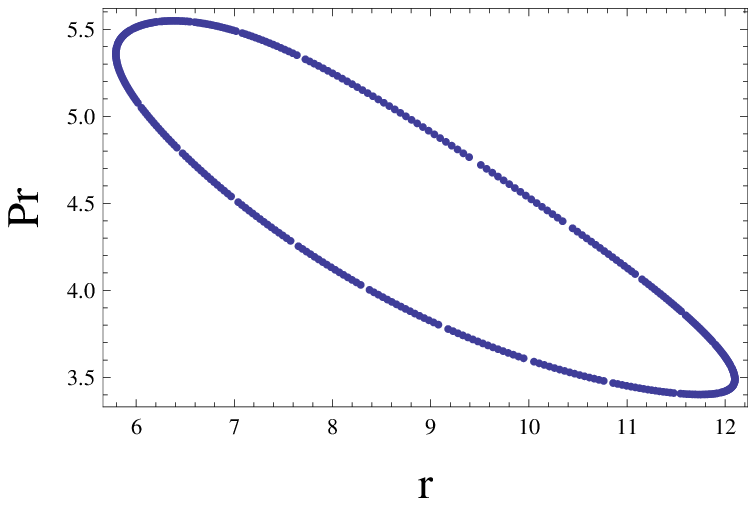}
\end{minipage}%
\begin{minipage}[b]{0.5\linewidth}
\includegraphics[width =2.7in,height=2in]{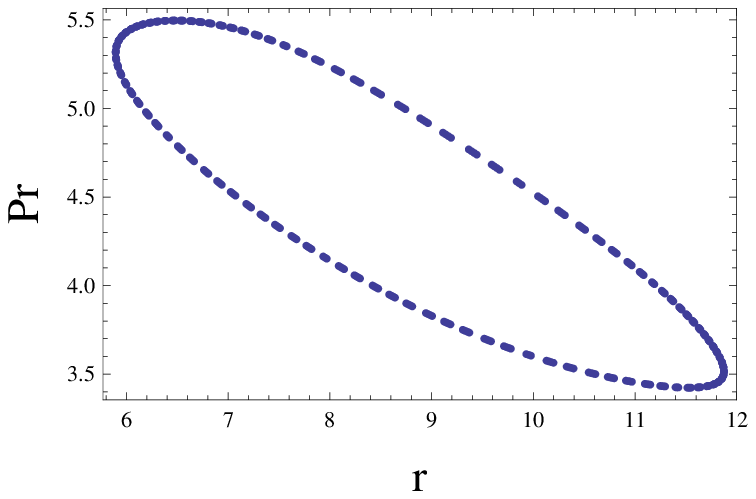}
\end{minipage}\quad
\begin{minipage}[b]{0.5\linewidth}
\includegraphics[width =2.7in,height=2in]{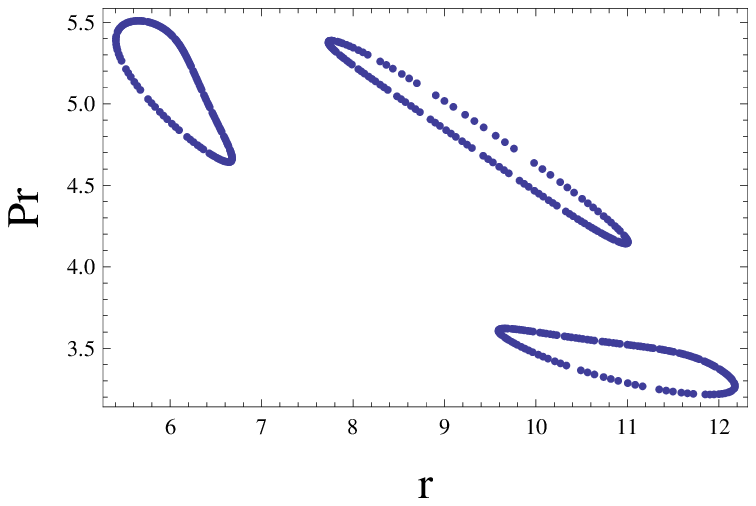}
\end{minipage}%
\begin{minipage}[b]{0.5\linewidth}
\includegraphics[width =2.7in,height=2in]{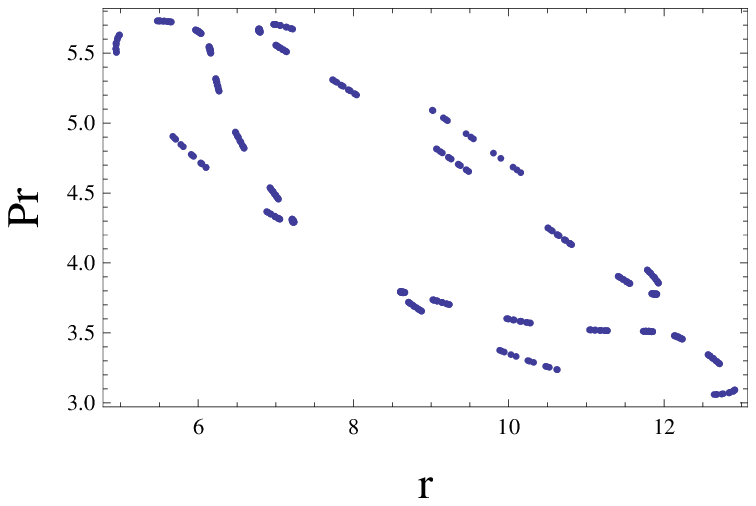}
\end{minipage}\quad
\caption{\label{fig:psecRNAdS} Plots depict the Poincare sections at $\theta(\tau)=0$ for the solutions $r(\tau)$ with increasing values of the charge $Q=(0.001,0.2,0.45)$ and $Q=0.4987$ (extremal case) of the RN-AdS black hole and initial conditions $E_n=12,P_r=3.6,~r=10$ and $\theta=0$.}
\end{figure}

In figure (\ref{fig:rvtRNAdS}) we plot the solution for $r(\tau)$ with initial conditions  $P_r(0)=3.6,~r(0)=10,~\theta(0)=0$, where $P_\theta(0)$ is given by the constraint (\ref{Hamil}). This plot is for different values of the black hole charge $Q=(0.001)$ and $Q=(0.45)$. In the same figure we also show the variation of ${\cal H}(\tau)$ with time which shows that the numeric constraint ${\cal H}<\delta$ is always satisfied for $\delta=10^{-10}$, confirming the robustness and accuracy of our analysis. Additionally, in figure (\ref{fig:psecRNAdS}) we have plotted the phase curve in the $(r,P_r)$ plane corresponding to the solutions $r(\tau)$ given in figure (\ref{fig:rvtRNAdS}). The phase curve seems to densely fill a given region of the  $(r,P_r)$ plane showing a rather complex pattern characteristic of deterministic chaotic systems.  In particular, we wish to show that the dynamical system of a ring string in the RN-AdS black hole background has a number of properties which indicate that it is hard to predict late time dynamics starting from two close neighboring points in phase space, corresponding to different initial conditions. To explore this further, we study the invariants, largest Lyapunove exponent $(LLE)$ and basin-boundary diagrams for our dynamical system.
\subsubsection*{Fixed points}
In order to analyze the late time dynamics of a given system it is essential to find out various possible positive invariant sets such as the fixed points, limit cycles etc. of the phase space. The invariant set has the property that every phase space curve starting at any point inside the invariant set remains within it forever. Thus, in particular our objective is to determine the fixed points for the phase space of the ring string in the RN-AdS black hole background, whose dynamics is defined by the system of equations $(3)$. To this end, we first set $(\dot{r}=0,\dot{P_r}=0,\dot{\theta}=0,\dot{P_{\theta}}=0)$ in the left hand side of the system of equations given in $(3)$. This leads us to the fixed point solution, $(\overset{*}{P_r}=0,\overset{*}{P_{\theta}}=0, \overset{*}{\theta}=(2k+1)\pi/2)$ with $\overset{*}{r}$ being given by $4 \overset{*}{r}^{3}/l^2+2 \overset{*}{r}-1=0$. This fixed point solution also satisfies the Hamiltonian constraint ${\cal H}=0$ given by the eq. (\ref{Hamil}) for all values of $k=\left(0,1,2\cdots\right)$. For a particular choice of $l=15$, we get $\overset{*}{r}=0.498896$ which is a fixed point located well behind the event horizon for all possible values of charge $Q$ of the black hole lying in the interval $(-0.5,0.5)$. The Jacobian for the system of equations $(3)$ at this fixed point with the condition (\ref{iniRNAdS}) gives the eigenvalues $\{34.0591,-34.0591,-1,1\}$ for $Q=0.001$; eigenvalues $\{-35.6451,35.6451,-1,1\}$ for $Q=0.2$; and eigenvalues $\{-60.4013,60.4013,-1,1\}$ for $Q=0.45$ respectively. The eigenvalues obtained for different values of the charge indicate that this fixed point is not an attractor (repellor) as all the eigenvalues are not negative (positive). 
There is also a fixed point given by $(\overset{*}{P_r}=0,\overset{*}{r}=\infty, \overset{*}{\theta}=k\pi,\overset{*}{P_{\theta}})$ for any arbitrary value of $\overset{*}{P_{\theta}}$, and can be determined by taking $r\rightarrow\infty$ limit in the system of equations given by $(3)$. For this fixed point if $\overset{*}{P_{\theta}}=0$ then for trajectories with $\overset{*}{P_{r}}>0$ ($\overset{*}{P_{\theta}}$), it acts as an attractor (repellor). Furthermore, if $\overset{*}{P_{\theta}}\neq 0$ then at some large but finite value of coordinate $r$ the angle $\theta$ changes which in turn leads to a large change in the momentum and $P_r$ changes sign, making the ring string to bounce back towards the black hole. Thus in a nutshell the invariant set corresponding to trajectories escaping to infinity has a zero measure which was also observed for the case of a ring string in Schwarschild AdS black hole background in \cite{Zayas:2010fs}.

\subsubsection*{Poincare sections}
The four dimensional $(r,P_r,\theta,P_{\theta})$  phase space for the ring string in RN-AdS black hole background can be reduced to a 3-dimensional space using the energy constraint (\ref{Hamil}) - in particular we set ${\cal H}=0$. This three dimensional manifold can be represented as a family of 2-dimensional concentric tori lying inside each other in a normal three dimensional space. For an integrable system any phase curve with any given initial condition will densely fill a 2-dimensional torus of the three dimensional energy manifold. Then a two dimensional plane in the three dimensional energy manifold intersecting the family of two dimensional tori in the transverse direction would constitute a co-dimension 2-Poincar\'{e} section. The Poincar\'{e} section may also be visualized as a family of concentric circles in the three dimensional energy manifold. For an integrable system any phase curve starting from a point (lying on a particular circle) on this plane (Poincar\'{e} section) would eventually cross it again after traversing the torus resulting in a new point (on the same circle), on the same plane. This is known as the Poincar\'{e} map. For a given Hamiltonian system the transition to chaos may be seen as the destruction of the circle in the Poincar\'{e} section, in other words the more diffused the Poincar\'{e} map for the system, the more chaotic it is. In figure (\ref{fig:psecRNAdS}) we have plotted the Poincar\'{e} sections at $\theta=0$ corresponding to the solutions $r(\tau)$ plotted in figure (\ref{fig:rvtRNAdS}) with increasing values of the charge $Q=(0.001,0.2,0.45)$ and $Q=0.4987$ (extremal value of the charge) from top to bottom respectively. We note that the system of a ring string in RN-AdS black hole background is weakly chaotic and becomes increasingly chaotic as we increase the charge of the black hole (i.e. as we approach extremality). 
\begin{figure}[H]
\centering
\begin{minipage}[b]{0.35\linewidth}
\includegraphics[width =2in,height=1.5in]{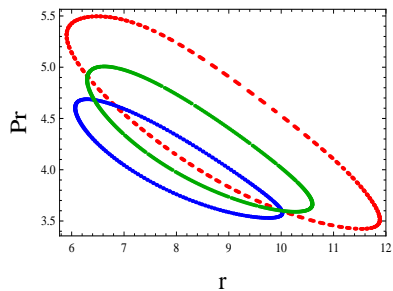}
\end{minipage}
\caption{\label{fig:psecRNAdS2} The red, green and the blue curves depict the Poincare sections at $\theta(\tau)=0$ for the solutions $r(\tau)$ with increasing values of the energy of the ring string $E_n=11,11.5,12$ respectively with initial conditions $Q=0.2,P_r=3.6,~r=10$ and $\theta=0$.}
\end{figure}
Moreover, in fig.(\ref{fig:psecRNAdS2}) we plot the Poincar\'{e} sections corresponding to the solutions $r(\tau)$ for fixed value of charge $Q=0.2$ with $P_r=3.6,~r=10, \theta=0$ and increasing values of the energy $E_n=11,11.5,12$. From the plots we observe that as we increase the energy of the ring string for fixed value of charge the area bounded by the Poincar\'{e} sections also increases. This indicates that the system becomes more chaotic with the increase in the energy of the ring string for a fixed value of the charge and other initial conditions. Although, we haven't shown the variation of the Poincar\'{e} sections for other fixed values of the RN-AdS black hole charge with increasing energy, a similar behavior is to be expected as presented for the case of the fixed value $Q=0.2$ of the charge of the RN-AdS black hole.

\subsubsection*{Largest Lyapunov exponent}
One of the characteristics of chaos is the dependence of the solutions on the initial conditions. This implies that if we consider a reference trajectory originating from a point of an invariant subset of the phase space, then we can always find a point in the vicinity of that invariant subset such that any trajectory starting from this point may diverge or converge from the reference trajectory at late times. Now if we consider two neighboring initial points in the phase space of a system denoted by $\xi_0$ (lying inside an invariant subset) and $\xi_0+\delta\xi_0$ (lying in the vicinity of the invariant subset ) then each of these will generate a trajectory in the phase space. These trajectories are functions of proper time interval, and so is the relative distance $(\Delta\xi(\xi_0,\tau))$ between them. The convergence of the trajectories at late times indicates existence of stable fixed point (attractor) in the system. However, if they happen to diverge from each other then it would indicate the presence of an unstable fixed point (repellor) in the system. Thus it becomes essential to study the mean exponential rate of divergence of the distance between two neighboring trajectories as a function of proper time which may be given as follows,

\begin{equation}
\lambda(\tau)=\left(\lim_{\Delta\xi_0\to 0}  \frac{1}{\tau}\ln \frac{\Delta\xi(\xi_0,\tau)}{\Delta\xi(\xi_0,0)} \right),\label{lleeq}
\end{equation}

\begin{figure}[H]
\centering
\begin{minipage}[b]{0.5\linewidth}
\includegraphics[width =2.7in,height=2in]{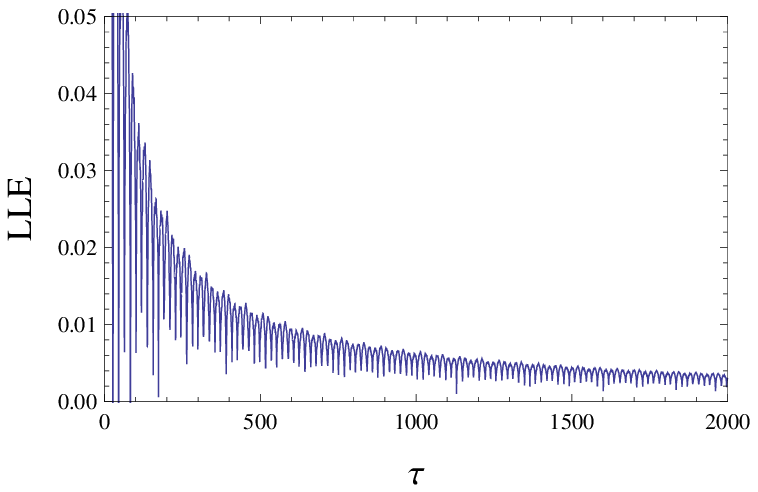}
\end{minipage}%
\begin{minipage}[b]{0.5\linewidth}
\includegraphics[width =2.7in,height=2in]{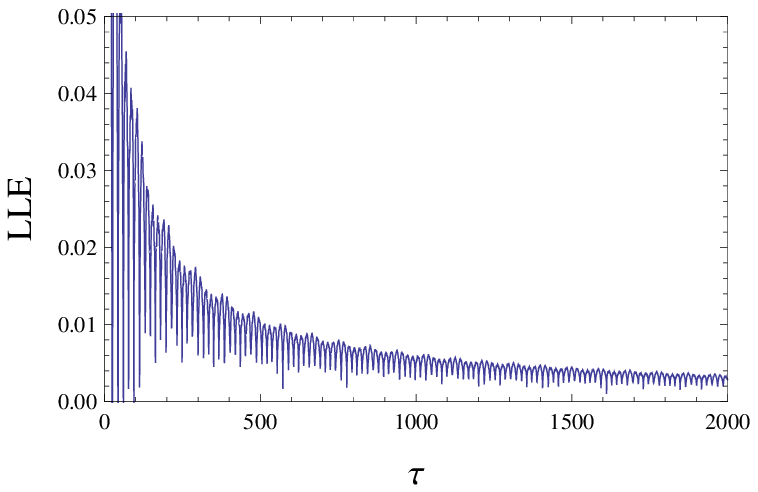}
\end{minipage}\quad
\begin{minipage}[b]{0.5\linewidth}
\includegraphics[width =2.7in,height=2in]{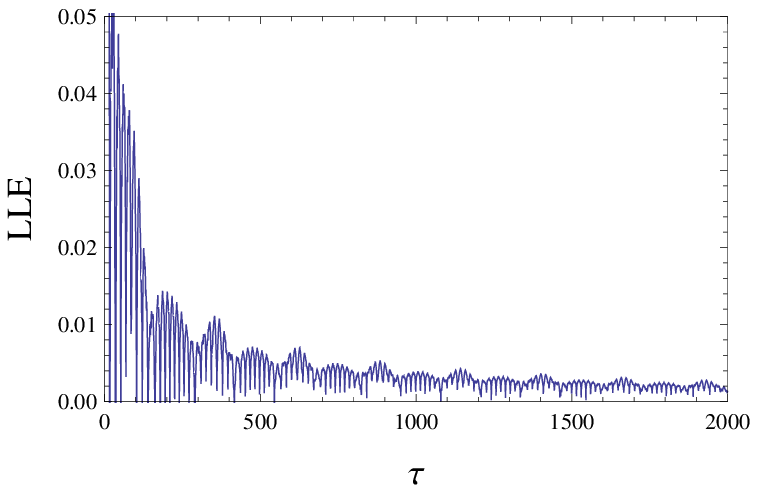}
\end{minipage}%
\begin{minipage}[b]{0.5\linewidth}
\includegraphics[width =2.7in,height=2in]{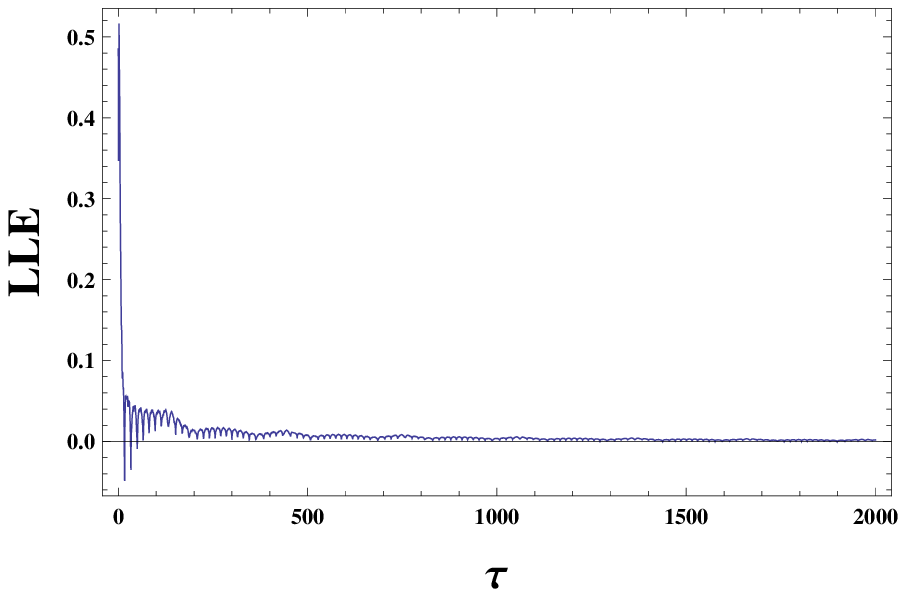}
\end{minipage}\quad
\caption{\label{fig:lleRNAdS} Plots  from top to bottom depict the largest Lyapunov exponent $(LLE)$ as a function of proper time $\tau$ for the solutions $r(\tau)$ with increasing values of the charge $Q=(0.001,0.2,0.45)$ and $Q=0.4987$ (extremal case)  of the RN-AdS black hole and the energy $E_n=12$ of the ring string as given in fig.(\ref{fig:rvtRNAdS}). }
\end{figure}
where $\lambda(\tau)$ is the {\it Largest Lyapunov Exponent} $(LLE)$. In figure (\ref{fig:lleRNAdS}) we have plotted the $LLE$ as a fuction of proper time corresponding to the solutions $r(\tau)$ plotted in figure (\ref{fig:rvtRNAdS}) - for increasing values of the charge $Q=(0.001,0.2,0.45,0.4987)$. The asymptotic value of this $LLE$  obtained at large times is known as the {\it characteristic Lyapunov Exponent}.  

In table (\ref{table:lleRnAdS}), we show values of the characteristic Lyapunov Exponent for various values of the charge $Q=(0.001,0.2,0.45)$ and $Q=0.4987$ (Extremal case) of the black hole. From the table we observe that the small asymptotic value of $LLE$ decreases further as we increase the charge of the black hole indicating a stronger rate of divergence of the distance between two neighboring trajectories. This leads us to conclude that our system is weakly chaotic due to small values of $LLE$ obtained at late times, and it also indicates that the system becomes more chaotic as the black hole charge increases. 
\begin{table}[H]
\centering
\begin{tabular}{|c|c|c|c|}
\hline 
\textbf{Charge of the black hole} &\multicolumn{2}{|c|}{$LLE$} \tabularnewline
\hline 
 & $E_n$=11 & $E_n$=12 \tabularnewline
 \hline
0.001 &0.001952192190361 &  0.002991542833306 \tabularnewline
\hline
0.2 & 0.001022679983635 & 0.002866314558667\tabularnewline
\hline
0.45 & 0.001892891218718 & 0.001325491001851 \tabularnewline
\hline
0.4987 & 0.002278089458435 & 0.001878828674668 \tabularnewline
\hline
\end{tabular}
\caption{\label{table:lleRnAdS} Table showing the values of the characteristic Lyapunov Exponent for various values of the charge $Q=(0.001,0.2,0.45)$ and $Q=0.4987$ (extremal case) of the RN-AdS black hole and energies $E_n=11,12$ of the ring string.}
\end{table}

\subsubsection*{Basins of Attraction}
\begin{figure}[H]
\centering
\begin{minipage}[b]{0.35\linewidth}
\includegraphics[width =2in,height=1.5in]{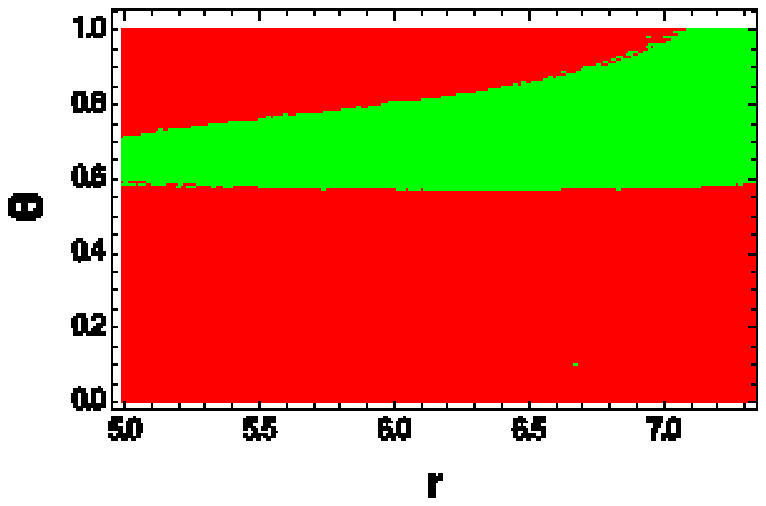}
\end{minipage}%
\begin{minipage}[b]{0.35\linewidth}
\includegraphics[width =2in,height=1.5in]{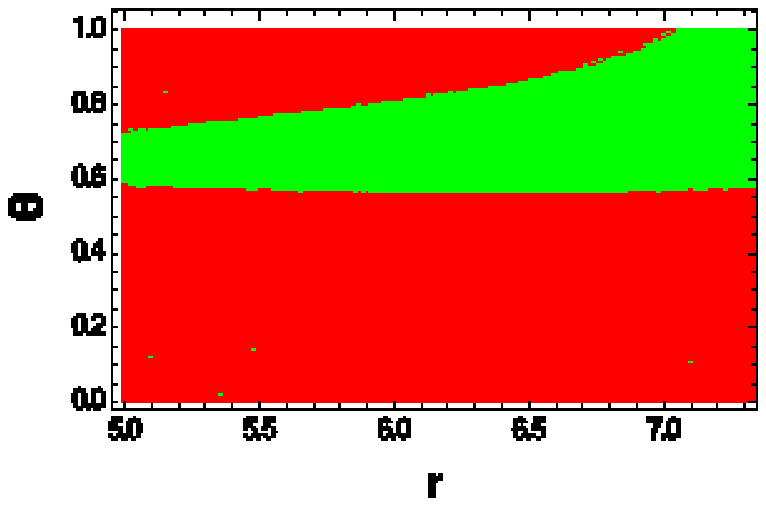}
\end{minipage}%
\begin{minipage}[b]{0.35\linewidth}
\includegraphics[width =2in,height=1.5in]{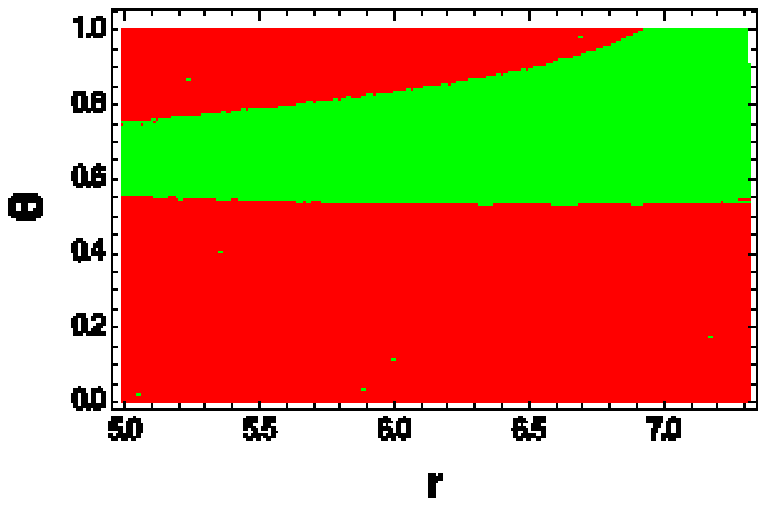}
\end{minipage}\quad
\caption{\label{fig:BasinAdSRN} Basin boundaries for the dynamics of the ring string in RN-AdS black hole background, plotted for the initial conditions  $E_n=12$, $l=15$ , $P_r(\tau)=0$ with $r\in[5,7.5]$ on the horizontal axis and $\theta\in[0,1]$ on the vertical axis. The graphs from left to right correspond to the values  $Q=0.001,0.02,0.045$ of the black hole charge respectively}
\end{figure}

For the ring string dynamics in RN-AdS black hole background, we once again adopt the basin-boundary method for coloring the two dimensional slice $(r,\theta)$ of the four dimensional space $(r,P_r,\theta, P_\theta)$ of initial conditions. The two dimensional slice of the initial conditions in $(r,\theta)$ is obtained by considering the constraint $P_r=0$, which in purpose fixes the value of $(P_\theta)$ in terms of $(r,\theta)$ , energy $(E_n)$ of the string, AdS scale $(l)$, and the charge $(Q)$ of the black hole. The figure (\ref{fig:BasinAdSRN}) depicts the basins of attraction of the motion of the ring string for different values of charge $(Q=0.001,0.2,0.45)$ of the RN-AdS black hole. As usual, we have colored the $(r,\theta)$ slice of initial conditions according to different asymptotic modes of motion that the ring string demonstrates in the RN-AdS black hole background. The red colored region in basin boundary diagrams given in fig. (\ref{fig:BasinAdSRN}) corresponds to the case when the ring string passes beyond $(r(\tau)<r_h)$ the radius of horizon $(r_h)$ and is captured by it. The green colored region depicts the case when the ring string is scattered by the black hole before escaping to infinity $(r\rightarrow\infty)$. In this case due to numerical reasons, instead of infinity the condition for escape is considered to be at some large but finite value of $r$ which corresponds to $r\geq 100 r_h$. Here too, the string may cross this cutoff in one direction and return back, however this will lead to incorrect coloring for very few points which follows from the asymptotic behavior of the Hamiltonian given by eq.(\ref{Hamil}). 
\begin{figure}[H]
\centering
\begin{minipage}[b]{0.35\linewidth}
\includegraphics[width =2in,height=1.5in]{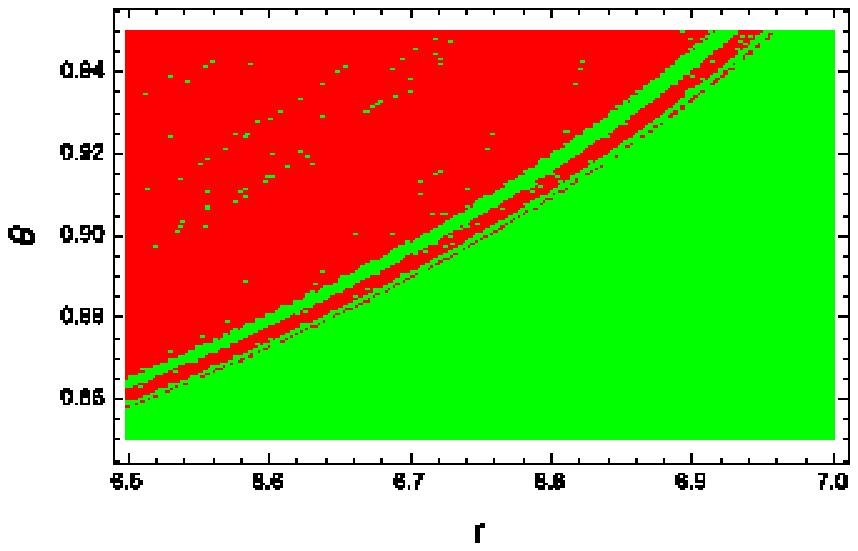}
\end{minipage}%
\begin{minipage}[b]{0.35\linewidth}
\includegraphics[width =2in,height=1.5in]{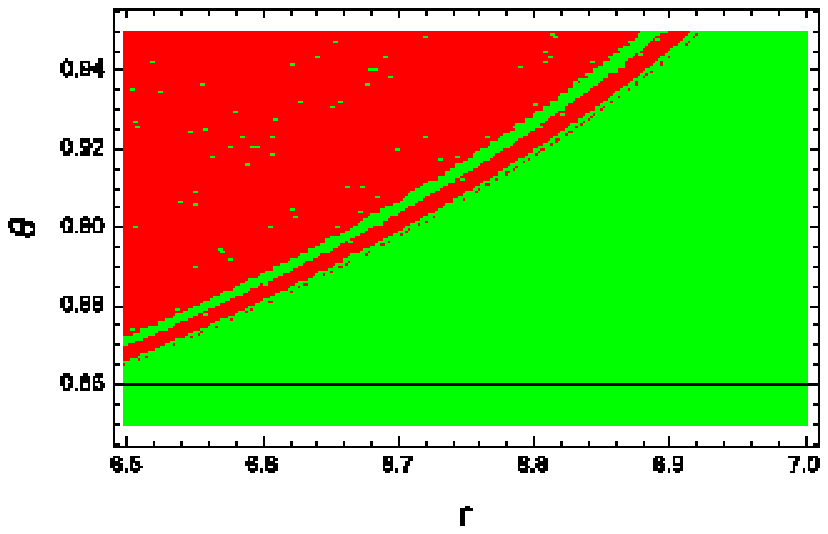}
\end{minipage}%
\begin{minipage}[b]{0.35\linewidth}
\includegraphics[width =2in,height=1.5in]{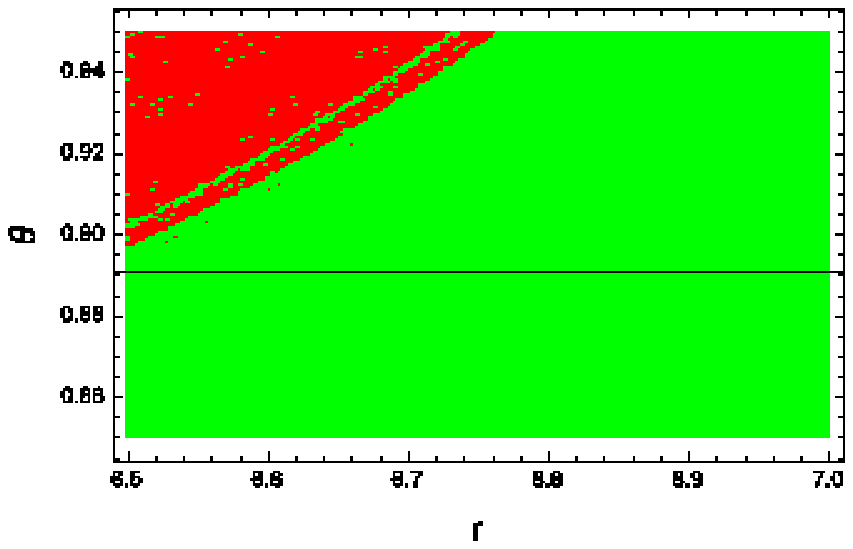}
\end{minipage}\quad
\caption{\label{fig:BasinAdSRN1} A zoom in of the Basin boundaries for the dynamics of the ring string in RN-AdS black hole background, plotted for the initial conditions $E_n=12$, $l=15$ , $P_r(\tau)=0$ with $r\in[6.5,7]$ on the horizontal axis and $\theta\in[0.835,0.955]$ on the vertical axis. The graphs from left to right correspond to the values  $Q=0.001,0.02,0.045$  of the black hole charge respectively.}\end{figure}
From the basin-boundary diagrams shown in figure (\ref{fig:BasinAdSRN}) it is clear that there are some well defined red and green regions which are separated from each other by boundaries which look fuzzy. For precision, we magnify a particular bounded region in figure (\ref{fig:BasinAdSRN1}) with the $r$- coordinate lying in the range $[6.5,7]$ and $\theta$-coordinate within the range $[0.835,0.955]$. From the magnified region of the basin boundary diagrams we observe that the red region tends to disappear as we increase the charge of the black hole. This implies that the tendency of the ring string getting back scattered by the back hole increases with increasing values of the black hole charge. The magnification of the basin-boundary region also reveals a fractal structure characteristic to the systems following deterministic chaos. This in turn indicates that the dynamics of the ring string in RN-AdS black hole background is chaotic, and becomes more so with increasing values of the black hole charge - as elucidated earlier by studying the Poincar\'{e} sections and the Largest Lyapunov exponents for the system. 


\section{Summary and Conclusions}
In summary we have investigated the dynamics and phase space behavior of a ring string in both - asymptotically flat and the asymptotically AdS, Reissner-Nordstrom black hole backgrounds. We numerically solved the set of nonlinear coupled differential equations governing the string dynamics. 

In the case of a ring string in asymptotically flat charged black hole background, it is observed that the dynamical system follows three dominant asymptotic modes involving either falling in to the black hole, escaping to infinity, or escaping to infinity via backscattering. Additionally, the ring string can also follow an infinite set of unstable periodic orbits. In order to determine the degree of chaos in the system, we studied the basin boundary diagrams by coloring the two dimensional slice $(r,\theta)$ of the four dimensional phase space $(r,P_r,\theta, P_\theta)$ of initial conditions depicting different modes of motion. From the basin diagrams it was observed that the tendency of the ring string escaping to infinity decreased with increasing values of the charge of the black hole. Remarkably, the magnification of the basin-boundary region revealed a fractal structure characteristic of deterministic chaos showcasing that the dynamics of the ring string in the asymptotically flat RN black hole background is indeed chaotic. 
 
In the case of a charged black hole in asymptotically AdS background the behavior is quiet different - it involves either falling in to the black hole,  oscillating a number of times around the black hole before collapsing into it, or escaping to infinity after completing a certain number of oscillations. We analyzed this dynamical system by studying the invariant set (fixed points) of the system, largest Lyapunov exponent $(LLE)$, and the usual basin-boundary diagrams. It was observed that the invariant set included a fixed point located well behind the event horizon for all possible values of charge $Q$ of the black hole without any possibility of it being an attractor or a repellor. However, analysis of the Poincar\'{e} sections for this system further revealed that the map gets more diffused with increase in the charge of the black hole. This indicates that the dynamics of the ring string in RN-AdS black hole background is weakly chaotic and the degree of chaos increases along with charge of the black hole (i.e. upon approaching the extremal limit). The basin-boundary diagrams for the system also point towards the same result. Additionally, we also conclude that the tendency of the ring string getting back scattered by the back hole increases with the increasing values of the charge of the black hole. Lastly, we examined the largest Lyapunov exponent $(LLE)$ for this system which corresponds to the mean exponential rate of divergence of the distance between two neighboring trajectories as a function of proper time. We observed that the small asymptotic value of $LLE$ decreased further as the charge of the black hole is increased, indicating a stronger rate of divergence of the distance between two neighboring trajectories. Therefore, all our investigations regarding a ring string in the charged black hole background, with asymptotically AdS boundary conditions, point to a weakly chaotic system characterized by small values of $LLE$ obtained at late times - with the feature that the system becomes increasingly chaotic as the charge is increased towards extremality. Needless to emphasize, it would also help to understand such string dynamics from a holographic perspective, which we hope to address in future.

\section{Acknowledgments}
This work of one of the authors, Pankaj Chaturvedi is supported by the Grant No. 09/092(0846)/2012-EMR-I, from the Council of Scientific and Industrial Research (CSIR), India.
\bibliographystyle{unsrt}
\bibliography{mybib}

\begin{thebibliography}{10}

\bibitem{Mandal:2002fs}
Gautam Mandal, Nemani~V. Suryanarayana, and Spenta~R. Wadia.
\newblock {Aspects of semiclassical strings in AdS(5)}.
\newblock {\em Phys. Lett.}, B543:81--88, 2002.

\bibitem{Bena:2003wd}
Iosif Bena, Joseph Polchinski, and Radu Roiban.
\newblock {Hidden symmetries of the $AdS_5 \times S^5$ superstring}.
\newblock {\em Phys. Rev.}, D69:046002, 2004.

\bibitem{Beisert:2010jr}
Niklas Beisert et~al.
\newblock {Review of AdS/CFT Integrability: An Overview}.
\newblock {\em Lett. Math. Phys.}, 99:3--32, 2012.

\bibitem{Giataganas:2013dha}
Dimitrios Giataganas, Leopoldo~A. Pando~Zayas, and Konstantinos Zoubos.
\newblock {On Marginal Deformations and Non-Integrability}.
\newblock {\em JHEP}, 01:129, 2014.

\bibitem{Giataganas:2014hma}
Dimitrios Giataganas and Konstadinos Sfetsos.
\newblock {Non-integrability in non-relativistic theories}.
\newblock {\em JHEP}, 06:018, 2014.

\bibitem{Basu:2011di}
Pallab Basu and Leopoldo~A. Pando~Zayas.
\newblock {Chaos rules out integrability of strings on AdS$_5 \times T^{1,1}$}.
\newblock {\em Phys. Lett.}, B700:243--248, 2011.

\bibitem{Basu:2011fw}
Pallab Basu and Leopoldo~A. Pando~Zayas.
\newblock {Analytic Non-integrability in String Theory}.
\newblock {\em Phys. Rev.}, D84:046006, 2011.

\bibitem{Basu:2012ae}
Pallab Basu, Diptarka Das, Archisman Ghosh, and Leopoldo~A. Pando~Zayas.
\newblock {Chaos around Holographic Regge Trajectories}.
\newblock {\em JHEP}, 05:077, 2012.

\bibitem{Stepanchuk:2012xi}
A.~Stepanchuk and A.~A. Tseytlin.
\newblock {On (non)integrability of classical strings in p-brane backgrounds}.
\newblock {\em J. Phys.}, A46:125401, 2013.

\bibitem{Chervonyi:2013eja}
Yuri Chervonyi and Oleg Lunin.
\newblock {(Non)-Integrability of Geodesics in D-brane Backgrounds}.
\newblock {\em JHEP}, 02:061, 2014.

\bibitem{Bai:2014wpa}
Xiaojian Bai, Bum-Hoon Lee, Taeyoon Moon, and Junde Chen.
\newblock {Chaos in Lifshitz Spacetimes}.
\newblock {\em J. Korean Phys. Soc.}, 68(5):639--644, 2016.

\bibitem{Ma:2014aha}
Da-Zhu Ma, Jian-Pin Wu, and Jifang Zhang.
\newblock {Chaos from the ring string in a Gauss-Bonnet black hole in AdS5
  space}.
\newblock {\em Phys. Rev.}, D89(8):086011, 2014.

\bibitem{Asano:2015qwa}
Yuhma Asano, Daisuke Kawai, Hideki Kyono, and Kentaroh Yoshida.
\newblock {Chaotic strings in a near Penrose limit of AdS$_{5} \times$
  T$^{1,1}$}.
\newblock {\em JHEP}, 08:060, 2015.

\bibitem{Panigrahi:2016zny}
Kamal~L. Panigrahi and Manoranjan Samal.
\newblock {Chaos in classical string dynamics in $\hat{\gamma}$ deformed $AdS_5
  \times T^{1,1}$}.
\newblock 2016.

\bibitem{Frolov:1999pj}
Andrei~V. Frolov and Arne~L. Larsen.
\newblock {Chaotic scattering and capture of strings by black hole}.
\newblock {\em Class. Quant. Grav.}, 16:3717--3724, 1999.

\bibitem{Zayas:2010fs}
Leopoldo~A. Pando~Zayas and Cesar~A. Terrero-Escalante.
\newblock {Chaos in the Gauge / Gravity Correspondence}.
\newblock {\em JHEP}, 09:094, 2010.

\end{thebibliography}

\end{document}